\documentclass{emulateapj} 

\usepackage{apjfonts}
\usepackage{lscape} 

\def \etal {et~al.~} 
\def \Yd   {\ifmmode \Upsilon_{\rm d} \else $\Upsilon_{\rm d}$ \fi} 
\def \YdR  {\ifmmode \Upsilon_{\rm d}^R \else $\Upsilon_{\rm d}^R$ \fi} 
\def \Rd   {\ifmmode R_{\rm d} \else $R_{\rm d}$ \fi} 
\def \rs   {\ifmmode r_{\rm s} \else $r_{\rm s}$ \fi} 
\def \rrm2  {\ifmmode r_{-2} \else $r_{-2}$ \fi} 
\def \ccm2  {\ifmmode c_{-2} \else $c_{-2}$ \fi} 
\def \cvir {\ifmmode c_{\rm vir} \else $c_{\rm vir}$ \fi } 
\def \Rvir {\ifmmode R_{\rm vir} \else $R_{\rm vir}$ \fi} 
\def \Vvir {\ifmmode V_{\rm vir} \else $V_{\rm vir}$ \fi} 
\def \Mvir {\ifmmode M_{\rm vir} \else $M_{\rm vir}$ \fi} 
\def \Deltavir {\ifmmode \Delta_{\rm vir} \else $\Delta_{\rm vir}$ \fi} 
\def \OmegaM {\ifmmode \Omega_{\rm M} \else $\Omega_{\rm M}$ \fi} 
 
\def \R200 {\ifmmode R_{200} \else $R_{200}$ \fi} 
\def \v200 {\ifmmode V_{200} \else $V_{200}$ \fi} 
\def \M200 {\ifmmode M_{200} \else $M_{200}$ \fi} 
 
\def \qd   {\ifmmode q_{\rm d} \else $q_{\rm d}$ \fi} 
\def \qh   {\ifmmode q_{\rm h} \else $q_{\rm h}$ \fi} 
 
\def \vmax {\ifmmode V_{\rm max} \else $V_{\rm max}$ \fi} 
\def \vmaxobs {\ifmmode V_{\rm max}^{\rm obs} \else $V_{\rm max}^{\rm obs}$\fi} 
\def \vdisk {\ifmmode V_{\rm disk} \else $V_{\rm disk}$ \fi} 
\def \vtot {\ifmmode V_{\rm tot} \else $V_{\rm tot}$ \fi} 
\def \vcirc {\ifmmode V_{\rm circ} \else $V_{\rm circ}$ \fi} 
\def \vrot {\ifmmode V_{\rm rot} \else $V_{\rm rot}$ \fi} 
 
\def \kms {\ifmmode \,\rm km\,s^{-1} \else $\,\rm km\,s^{-1} $ \fi } 
\def \kpc {\ifmmode {\rm kpc}      \else ${\rm kpc}$       \fi } 
\def \Msol {\ifmmode \rm M_{\odot} \else $\rm M_{\odot}$  \fi } 
\def \hMsol {\ifmmode h^{-1}\,\rm M_{\odot} \else $h^{-1}\,\rm M_{\odot}$ \fi} 
\def \Msolpc2{\ifmmode\rm M_{\odot}\,pc^{-2}\else $\rm M_{\odot}\,pc^{-2}$\fi} 
 
\def \chisq {\ifmmode \chi^2 \else $\chi^2$ \fi} 
\def \chisqr {\ifmmode \chi^2_{\rm r} \else $\chi^2_{\rm r}$ \fi} 
 
\def \spose#1{\hbox to 0pt{#1\hss}} 
\def\lta{\mathrel{\spose{\lower 3pt\hbox{$\sim$}}\raise 2.0pt\hbox{$<$}}} 
\def\gta{\mathrel{\spose{\lower 3pt\hbox{$\sim$}}\raise 2.0pt\hbox{$>$}}} 
              
\def\h{\rm h} 
\def\m{\rm m} 
 
\def\b{\rm b} 
\def\c{\rm c} 
\def\B{\rm B} 
 
\def \ion#1#2{#1{\footnotesize{#2}}\relax} 
\def \ha       {H$\alpha$} 
\def \hi       {\ion{H}{I}}

\shorttitle{MASS MODELING OF DISK GALAXIES} 
\shortauthors{DUTTON ET~AL.} 
\journalinfo{THE ASTROPHYSICAL JOURNAL, 619:218-242, 2005 January 20} 
\submitted{\it Recieved  2003 september 30;  accepted 2004 September
15}
 
\begin{document} 
 
\title{Mass Modeling  of Disk Galaxies:  Degeneracies, Constraints \\
 and Adiabatic Contraction}
 
\author{Aaron A. Dutton} 
\affil{Department of Physics \& Astronomy, University of British
  Columbia, 6224 Agricultural Road, Vancouver, \\ BC V6T 1Z1, Canada; and
Institute of Astronomy, Department of Physics, ETH Z\"{u}rich, \\
Scheuchzerstrasse 7, 8093 Z\"{u}rich, Switzerland; dutton@phys.ethz.ch} 

\author{St\'{e}phane  Courteau} \affil{Department of  Physics, Queen's
University, Kingston ON, K7L 3N6, Canada;
courteau@astro.queensu.ca}

\author{Roelof de Jong} \affil{Space Telescope Science Institute, 
3700 San Martin Dr., Baltimore, MD 21218, USA; dejong@stsci.edu} 

\and

\author{Claude Carignan} 
\affil{D\'{e}partment de Physique, Universit\'{e} de Montr\'{e}al, C.P.  6128, Station Centre-Ville, Montr\'{e}al, \\
QC H3C 3J7, Canada; carignan@astro.umontreal.ca\\}

\begin{abstract} 
This paper  addresses available constraints  on mass models  fitted to
rotation  curves.   Mass  models  of  disk  galaxies  have  well-known
degeneracies,  that  prevent a  unique  mass  decomposition. The  most
notable is due to the unknown value of the stellar mass-to-light ratio
(the disk-halo degeneracy); even with this known, degeneracies between
the   halo   parameters   themselves   may  prevent   an   unambiguous
determination of  the shape of  the dark halo profile,  which includes
the inner density slope of the  dark matter halo.  The latter is often
referred to as the ``cusp-core degeneracy.''
We explore constraints on the disk and halo parameters and apply these
to four mock  and six observed disk galaxies  with high resolution and
extended  rotation curves.  Our  full set  of constraints  consists of
mass-to-light ($M/L$) ratios  from stellar population synthesis models
based on  $B-R$ colors, constraints  on halo parameters  from $N$-body
simulations, and constraining the halo virial velocity to be less than
the maximum  observed velocity.  These constraints  are only partially
successful in lifting the cusp-core degeneracy.
The effect  of adiabatic  contraction of  the halo by  the disk  is to
steepen cores  into cusps and  reduce the best-fit  halo concentration
and $M/L$ values (often significantly).
We  also  discuss  the  effect  of disk  thickness,  halo  flattening,
distance  errors, and rotation  curve error  values on  mass modeling.
Increasing  the imposed  minimum rotation  curve error  from typically
low, underestimated  values to more realistic  estimates decreases the
$\chisq$  substantially and  makes distinguishing  between a  cuspy or
cored halo profile even more difficult.
In  spite   of  the   degeneracies  and  uncertainties   present,  our
constrained mass  modeling favors sub-maximal disks  (i.e., a dominant
halo)  at 2.2 disk  scale lengths,  with $\vdisk/\vtot\lta  0.6$. This
result holds for both the  un-barred and weakly barred galaxies in our
sample.
\end{abstract} 
 


\section{Introduction}\label{sec:intro} 
 
There has  been significant  debate recently about  the shape  of dark
matter  density  profiles,  especially  regarding their  inner  slope,
$\alpha$.\footnote{Values  of  $\alpha$ range  from  0  (core) to  1.5
(cuspy).  We define  the dark matter profile and  $\alpha$ in Equation
(4).}  Based on cosmological  $N$-body simulations (Navarro, Frenk, \&
White 1996;  Navarro, Frenk, \&  White 1997; hereafter NFW),  the dark
matter halo profile appears to be independent of mass and has an inner
logarithmic   slope  $\alpha=1$.    More  recent,   higher  resolution
simulations suggest  that the  density profiles do  not converge  to a
single  power-law at  small radii.   At the  smallest  resolved scales
($\simeq  0.5\%$ of the  virial radius)  profiles usually  have slopes
between 1 and  1.5 (Moore \etal 1999; Ghigna \etal  2000; Jing \& Suto
2000; Fukushige \&  Makino 2001; Klypin \etal 2001;  Power \etal 2003;
Navarro \etal 2004; Diemand \etal 2004).
 
At  large radii  all  simulations find  density  profiles with  slopes
$\alpha \simeq  -3$, which is  {\it inconsistent} with  the isothermal
($\rho \propto  r^{-2}$) profile. The  determination of the  dark halo
slope  based on  mapping  the  outer density  profile  of galaxies  is
difficult,  owing mainly to  a lack  of mass  tracers at  large radii.
Prada \etal (2003) find  that the line-of-sight velocity dispersion of
satellite galaxies declines with distance to the primary, in agreement
with a $\rho \propto r^{-3}$ density profile at large radii.
 
The  determination of  $\alpha$  based  on data  at  smaller radii  is
complicated by  the unknown value of the  stellar mass-to-light ratio,
$\Yd$.   This has  led to  dedicated analyses  on dwarf\footnote{Dwarf
spiral  galaxies are  usually  defined as  having  a maximum  rotation
velocity   $v_{\rm{max}}\,<\,100\,\rm{km\,s^{-1}}$   and/or  a   total
magnitude $M_{\B}  \ge -18$.}  and  low surface brightness\footnote{An
LSB galaxy  is usually defined as  a disk galaxy  with an extrapolated
central    disk   surface    brightness    $\mu_0^{\B}$   roughly    2
$\rm{mag}\,\rm{arcsec}^{-2}$ fainter than the typical value for HSB
 galaxies of  $\mu_0^{\B} =  21.65$ (Freeman
1970).} (LSB) galaxies  that are believed to be  dark matter dominated
at all radii (de Blok \& McGaugh 1997; Verheijen 1997; Swaters 1999).
 
It has been  suggested that rotation curves of  dwarf and LSB galaxies
rise less steeply than predicted by numerical simulations based on the
cold dark matter  (CDM) paradigm (Moore 1994; Flores  \& Primack 1994;
de  Blok \&  McGaugh 1997;  McGaugh  \& de  Blok 1998;  de Blok  \etal
2001a,b).
However, a number of observational uncertainties cast doubt over these
early  conflicting  claims.  These  include  beam  smearing for  \hi\,
rotation curves (Swaters  \etal 2000; van den Bosch  \etal 2000), high
inclination angles and \ha\,  long-slit alignment error (Swaters \etal
2003a),  and non-circular  motions  close to  the  center of  galaxies
(Swaters \etal 2003b).  Many  of these uncertainties can be quantified
or  eliminated by  measuring high-resolution  two-dimensional velocity
fields (Barnes, Sellwood, \&  Kosowsky 2004).  At optical wavelengths,
these   can  be   obtained  via   Fabry-Perot   interferometry  (e.g.,
Blais-Ouellette  \etal  1999) or  integral  field spectroscopy  (e.g.,
Andersen \& Bershady 2003; Courteau \etal 2003).
 
Despite a  low ratio of baryonic  to non-baryonic matter  in dwarf and
LSB  galaxies,  practical limitations  in  accurately determining  the
circular velocity  profile have prevented a  reliable determination of
the dark  matter density profile for those  galaxies. Furthermore, the
predictions of numerical simulations are weakest on the (small) scales
of dwarf and LSB galaxies.  By comparison, for high surface brightness
(HSB) galaxies  the kinematics is  easier to measure and  the expected
dark halos  can be better  resolved in numerical simulations,  but the
more  prominent   stellar  component  often  hinders   a  unique  mass
decomposition.
 
In principle, if the disk  mass-to-light ratio, $\Yd$, and the gaseous
mass distribution  are known, the  contribution from the dark  halo to
the  overall potential  can  be determined.   However, extracting  the
parametrized halo profile with  this procedure is complicated owing to
a  degeneracy between the  halo parameters  themselves (e.g.,  van den
Bosch \&  Swaters 2001).  Furthermore,  various evolutionary processes
may  alter the  dark  halo density  profile  from that  found in  dark
matter-only  simulations. The dissipation  of the  disk is  thought to
compress  the  dark halo  distribution  through adiabatic  contraction
(Blumenthal \etal 1986; Flores \etal 1993), while other processes such
as feedback,  mergers, spin segregation  (Maller \& Dekel  2002; Dekel
\etal  2003), pre-processing  of  dark  halos (Mo  \&  Mao 2003),  and
bar-driven dark halo evolution (Weinberg  \& Katz 2002) are thought to
lower the concentration and central cusp of dark matter halos.
 
In this  paper we  discuss and apply  mass modeling constraints  in an
attempt to break internal modeling degeneracies and thus determine the
best parameterization of the dark  halo. We present our mass models in
\S 2  and their degeneracies in  \S 3. The mass  model constraints are
presented in  \S 4.  We then  apply these constraints  to six galaxies
from Blais-Ouellette (2000).   The data are presented in  \S5, and the
models are applied to the data  in \S6.  The effects of rotation curve
errors, distance, disk thickness, and halo flattening are discussed in
\S7, and a summary is offered in \S 8.
 
Throughout this paper $r$ and $R$  refer to the radius from the galaxy
center      in     spherical     and      cylindrical     coordinates,
respectively. Whenever necessary, we also  adopt a value of the Hubble
constant\footnote{The current best estimate  of the Hubble constant is
  $H_0=72 \pm 8\,\rm{km}\,\rm{s}^{-1}\,\rm{Mpc}^{-1}$ ($HST$ $H_0$ Key
  Project; Freedman et al. 2001).}  
$H_0$ given by $h=H_0/100=0.7$.


\section{Mass Models}\label{sec:models} 
Our mass models consist of three main components for each disk galaxy:
a thick stellar disk (hereafter the ``disk''), an infinitesimally thin
gas disk (hereafter  the ``gas''), and an oblate  dark halo (hereafter
the  ``halo'').  In  general,  disk  galaxies may  also  have a  bulge
component,  although for simplicity  we limit  our analysis  to nearly
bulge-less systems.  Assuming that  the matter distribution is axially
symmetric and  in virial equilibrium,  the total circular  velocity is
given by
\begin{equation} 
  V_{\rm   circ}   =    \sqrt{V_{\rm{gas}}^2   +   V_{\rm{disk}}^2   +
 V_{\rm{halo}}^2},
\end{equation} 
at each radius $R$. Each of  the three components is described in more
detail below. We  compute the circular velocities of  the disk and gas
using formula A.17 of Casertano (1983). The best-fitting mass model is
determined  by  fitting  \vcirc  to  the  observed  circular  rotation
velocity,   $\vrot$,  by  minimizing   the  \chisq-statistic   with  a
non-linear optimization scheme.

\subsection{Stellar Disk} 
We model the disk with the following density profile (van der Kruit \&
Searle 1981):
\begin{equation} 
\rho_{\rm disk}(R,z) = \frac{\Sigma(R)\, {\rm{sech}}^2(z/z_0)}{2 z_0},
\end{equation} 
where $\Sigma(R)$ is the disk surface density profile and $z_0$ is the
vertical scale height. Unless otherwise stated, we compute $\Sigma(R)$
from the observed surface brightness profile.
 
The vertical scale  height is parameterized in terms  of the intrinsic
disk thickness,  $\qd \equiv z_0/\Rd$,  where $\Rd$ is the  disk scale
length.  Unless  otherwise stated,  we adopt $\qd=0.25$  (Kregel \etal
2002;  Bizyaev \&  Mitronova 2002).   We  explore the  effect of  disk
thickness in \S 7.

\subsection{Gaseous Disk} 
We model the gas disk with the following density profile:
\begin{equation} 
\rho_{\rm{gas}}(R,z) = \delta(z) \, \Sigma_{\rm HI}(R) / f_{\rm HI}
\end{equation} 
where  $\delta(z)$  is  the  Kronecker  delta  function,  $\Sigma_{\rm
HI}(R)$ is the  surface density of neutral hydrogen,  and $f_{\rm HI}$
is  the fraction of  gas in  \hi.  We  adopt $f_{\rm  HI}=0.75$ (e.g.,
Blais-Ouellette et  al.  2001); other  authors take $ 0.71  \le f_{\rm
HI} \le 0.77$, although the exact value is not critical.
 
Some spiral galaxies  show a central depression in  the \hi\, density,
likely due  to the gas being  present in a different  form (ionized or
molecular)  and/or  partially   or  completely  consumed  in  previous
episodes of star formation.  A central depression and hence a positive
radial density gradient result in  an outward radial force or negative
$V^2_{\rm  gas}$.   We represent  this  as  negative  velocity on  the
gaseous component of the rotation curve.

\begin{figure}[t] 
\begin{center}
\figurenum{1} 
\includegraphics[bb=  50 200 570 703, width=3.4in]{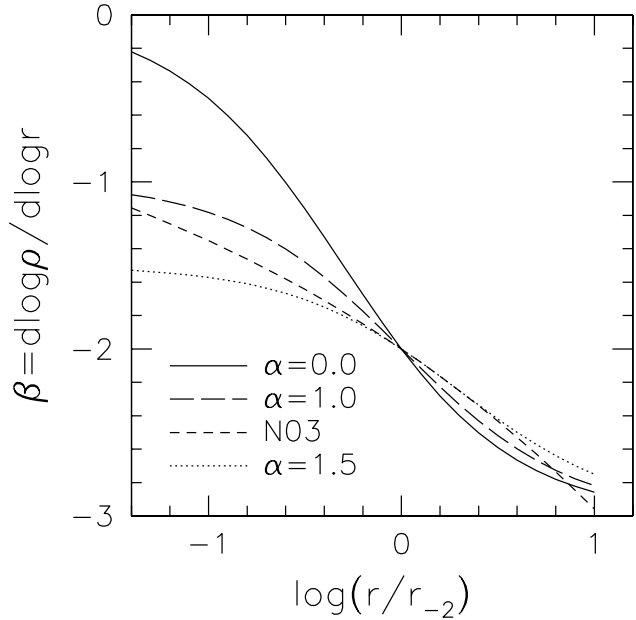}
\caption{Logarithmic slope of the 'ALP' halo density profiles for  
$\alpha=0,1$, and $1.5$ and $c=10$ plotted from $0.4\,\%\,r_{200}\le r
\le r_{200}$.  The highest resolution $N$-body simulations can resolve
the  density profile over  this range.   Typical density  profiles lie
between the  $\alpha=1$ and  $\alpha=1.5$ lines.  For  comparison with
the  'ALP' parameterization we  show the  fitting function  of Navarro
\etal 2004.}
\vspace{-0.25in}
\end{center}
\end{figure} 

\begin{figure}[b] 
\figurenum{2}
\includegraphics[bb=  50 200 570 703, width=3.4in]{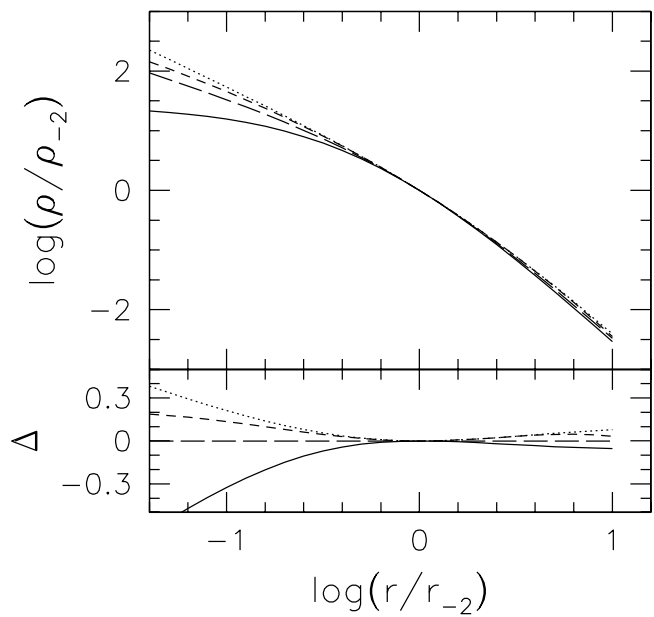}
\caption{Density  profiles  for  halos  in Fig.~1  normalized  to  the
  density at the scale radius, $\rho_{-2}$. The bottom panel shows the
  differences with respect to the $\alpha=1$ profile.}
\vspace{-0.3in}
\end{figure} 

\begin{figure}[t] 
\figurenum{3}
\includegraphics[bb=  50 200 570 703, width=3.4in]{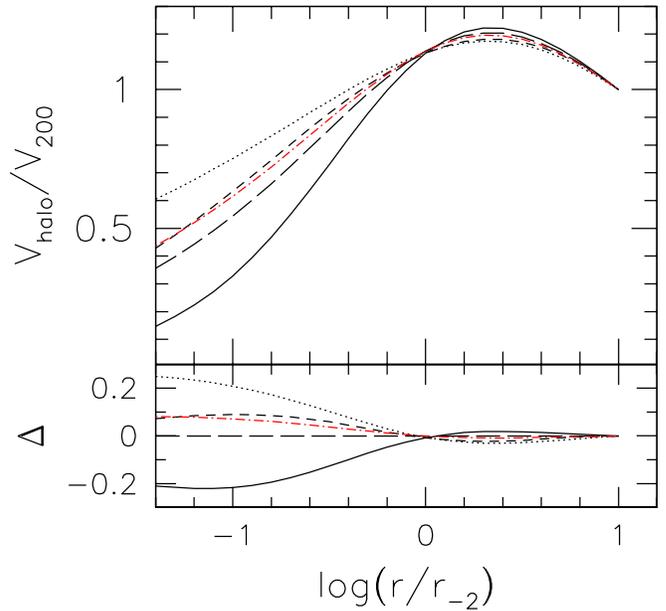}
\caption{Circular velocity profiles for halos in Fig.~1 
normalized to the virial  velocity, $V_{200}$.  The bottom panel shows
the differences with respect to the $\alpha=1$ profile.  Note that the
differences between  the profiles are most conspicuous  for radii less
than $\sim 0.5\,r_{-2}$. Also shown  is a halo with $\alpha=1.2$ ({\it
red dot-dashed line}), which is effectively indistinguishable from the
Navarro \etal (2004) fitting function.}
\end{figure}

\subsection{Dark Halo} 
To  take account  of the  uncertainties  in the  predicted inner  halo
density  profiles and  to allow  for halos  with flat  central density
profiles while preserving the $\rho\propto r^{-3}$ dependence at large
radii, we  use the following  density profile (hereafter  ALP profile;
Kravtsov \etal 1998):
\begin{equation} 
\rho_{\rm  halo}(r)= \frac{\rho_0}  {(r/r_{\rm s})^{\alpha}(1+r/r_{\rm
s})^{3-\alpha}}.
\end{equation} 
This density profile  has an inner logarithmic slope  of $-\alpha$ and
an outer logarithmic  slope of -3. For $\alpha=1$  this reduces to the
NFW profile, and at the scale  radius, $\rs$, the slope of the density
profile is -2.
 
However,  for  different  values  of $\alpha$,  $\rs$  corresponds  to
different  density slopes.  To  enable an  easier comparison  of scale
radii, we replace  $\rs$ with $r_{-2}$, the radius  where the slope of
the  density profile  is  $-2$.  With  the  conversion $r_{-2}  \equiv
(2-\alpha)\rs$.
 
Figures 1, 2, and 3 show the logarithmic density slopes, density
profiles, and circular velocity  profiles for halos with $\alpha=0,1,$
and $1.5$ and the fitting  function from Navarro \etal 2004 (which has
effectively $\alpha\simeq1.2$).

\subsubsection{Oblate/Prolate Density Profiles} 
Typically rotation  curve analyses assume  a spherical dark  halo even
though   CDM  simulations  suggest   triaxial  shapes   for  collapsed
structures, with  typical axis ratios  $c/a=0.5-0.7$ and $b/a=0.7-0.9$
(Dubinski \& Carlberg 1991; Jing  \& Suto 2002; Tinker \& Ryden 2002).
However,  the dissipative  infall of  gas in  non-baryonic  dark halos
suppresses triaxial structures, leading  to halos with an oblate shape
(Katz \&  Gunn 1991; Dubinski 1994; although  further investigation is
needed to quantify this effect).
 
This tentative  conclusion agrees with a variety  of observations that
find axially  symmetric disks,  with eccentricity $e  < 0.05$  ($b/a >
0.9987$; Combes  2002 and references  therein). The flattening  of the
halo  is not easily  measured, but  various techniques,  including the
flaring of \hi\, disks, polar  rings around spiral galaxies, and X-ray
isophotes of  elliptical galaxies, suggest  oblate halos with  an axis
ratio, $q=c/a$, ranging from 0.1 to 0.9 (Combes 2002).
 
Thus, we are compelled to  study the effects of axially symmetric dark
halos ($b/a=1$) in our mass models. We generalize the density profiles
to   the   family  of   axially   symmetric   ellipsoids  by   setting
$\rho(r)=\rho(m)$, where $m^2=R^2+z^2/q^2$.

\subsubsection{$c-V_{200}$ parameterization} 
We choose to parameterize the density profile by the circular velocity
at the virial radius,  $\Vvir$, and the concentration parameter, \ccm2
= \Rvir/$\rrm2 $. Here \Rvir is  the virial radius in the $z=0$ plane.
We choose  to define the virial  radius, $\Rvir$, as  the radius where
the mean density of the halo is \Deltavir times the critical density,
\begin{equation} 
\bar{\rho}(m_{\rm           vir})           =           \frac{M(m_{\rm
vir})}{\frac{4}{3}\pi\,q\,m_{\rm  vir}^3} =\Delta_{\rm vir}\,\rho_{\rm
crit},
\end{equation} 
where  $\rho_{\rm   crit}=\frac{3\,H^2}{8\,\pi\,G}$  is  the  critical
density of  the universe. With  these definitions $\m_{\rm  vir}$, and
hence  $\Rvir$,  will  be  invariant  under changes  of  $q$.   Unless
otherwise  stated, we  adopt \Deltavir  = 200,  although  in currently
favored $\Lambda$CDM cosmologies, at  redshift zero, the virial radius
occurs  at \Deltavir  =  337  \OmegaM $\simeq$  100  (Bryan \&  Norman
1998). The  choice of $\Delta_{\rm  vir}$ does not affect  the density
profile, but the virial radius changes by a factor of $\sim 1.3$.
 
With these definitions  and using Equation 2.91 in  Binney \& Tremaine
(1987) for   the   computation  of   $V_{\rm   circ}$,  the   velocity
contribution of  the halo specified by  $V_{200}$, $c_{-2}$, $\alpha$,
and $q$ is given by
\begin{equation} 
V_{\rm  halo}^2(x,z=0)  =V_{200}^2  \frac{\mu(x,\alpha,q)/x}{\mu(\ccm2
,\alpha,q)/\ccm2 }, \;\;x=R/r_{-2}
\end{equation} 
where
\begin{equation} 
\mu(x,\alpha,q)=\int_0^x   \frac{y^{2-\alpha}[1-(2-\alpha)y]^{\alpha-3}
} { \sqrt{1-(1-q^2) y^2/x^2} }\,dy.
\end{equation} 
 
 With the  above definitions we  can express the  relationship between
 \Rvir and \Vvir as
\begin{equation} 
\left(\frac{V_{\rm             vir}}{R_{\rm             vir}}\right)^2
=h^2\left(\frac{\Delta_{\rm                           vir}}{200}\right)
\frac{\mu(c_{-2},\alpha,q)}{\mu(c_{-2},\alpha,1)}
\end{equation} 
with   \Vvir  and   \Rvir   in  \kms   and   kpc,  respectively,   and
$h=H_0/100=0.7$.
 
Fig. 4 shows  the effect of $q$ on the circular  velocity of the halo,
normalized by  $\Vvir$. Note that for  a given $\Rvir$,  $\ccm2 $, and
$\alpha$,  \Vvir increases  as  $q$ decreases.   Oblate ($q<1$)  halos
result  in higher  circular  velocities, especially  near the  center,
while prolate ($q>1$) halos result in lower circular velocities.

\begin{figure}[t] 
\begin{center}
\figurenum{4} 
\includegraphics[bb=  50 200 570 703, width=3.3in]{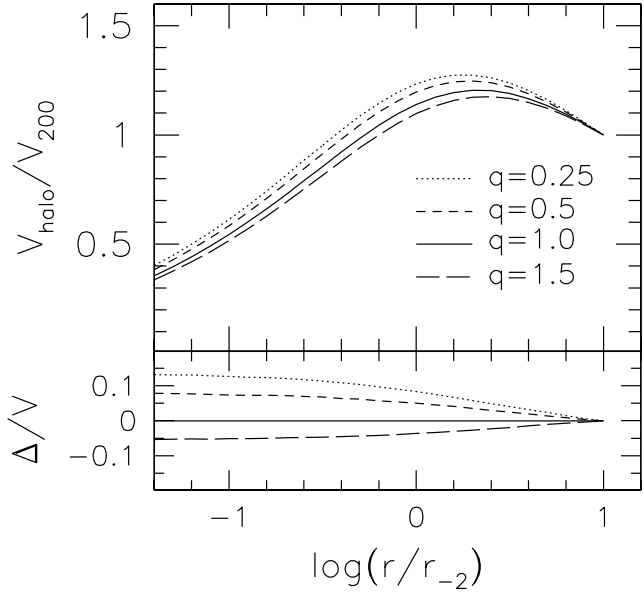}
\caption{Effect of halo flattening $q$ on circular velocity profiles for  
$\alpha=1$ halos with $c=10$. 
The bottom panel shows the fractional differences with respect to the $q=1$  
case. Note that these differences increase with decreasing radii.} 
\vspace{-0.2in}
\end{center}
\end{figure}

\subsubsection{Adiabatic Contraction} 
We model  the response of the  dissipation-less halo to  the infall of
the  dissipational  baryons  as  they  cool and  settle  into  a  disk
following  Blumenthal  \etal (1986)  and  Flores  \etal (1993).   This
assumes  that  the  collapse  of  the  baryons  is  slow,  the  matter
distribution is spherically symmetric,  and particles move on circular
orbits. Then  the adiabatic invariant is simply  $rM(r)$, where $M(r)$
is the mass enclosed by  radius $r$. With the further assumptions that
the dark matter particles do not cross orbits,
$M_{\rm{DM}}(r_{\rm{f}}) = M_{\rm{DM}}(r_{\rm{i}})$,
where $r_{\rm{i}}$ and $r_{\rm{f}}$ are the initial and final radii of
the disk, respectively, and that  the baryons are initially mixed with
the dark matter with a baryon fraction
$f_{\rm{B}} = M_{\rm{B}} / (M_{\rm{B}}+M_{\rm{DM}} )$,
then    given    the   {\it    initial}    dark   halo    distribution
$M_{\rm{DM}}(r_{\rm{i}})$ and final baryonic mass distribution $M_{\rm
B}(r_{\rm  f})$, the final  radius, $r_{\rm{f}}$,  can be  obtained by
solving
\begin{equation} 
  r_{\rm{f}}  [{M_{\rm{B}}(r_{\rm{f}})  + M_{\rm{DM}}(r_{\rm{i}})}]  =
r_{\rm{i}} {M_{\rm{DM}}(r_{\rm{i}})/(1-f_{\rm{B}})}.
\end{equation} 
 
We obtain the baryonic mass from  the observations of stars and gas in
the disk  (\S2.1), for  a given $\Upsilon_{\rm  d}$ and  distance, and
assume that  the fraction of baryons  in the halo  is negligible.  For
the mass of the halo we assume the virial mass, $M_{\rm vir}$.
 
The  effect  of adiabatic  contraction  on  the  density and  circular
velocity distributions  can be  quite substantial. We  illustrate this
effect in Figure~5.   This shows the circular velocity  of the initial
and  final halo and  final disk.   Here the  disk is  exponential with
$R_d=2$kpc,     $\mu_0^R=20~$mag~arcsec$^{-2}$     ($R$-band),     and
$\Upsilon_{\rm   d}^R=1.0$  and   0.25.   The   effect   of  adiabatic
contraction  is largest  for halos  with  low values  of $\alpha$  and
$\c_{-2}$, such that {\it halos with initial cores end up with cusps}.
For very cuspy  halos, the halo can expand in the  very center, as the
final  baryonic mass  within  $r_{\rm  i}$ is  less  than the  initial
baryonic mass within $r_{\rm i}$.  Note that although the mass is more
centrally  concentrated   after  adiabatic  contraction,   the  formal
concentration, $c_{-2}$, can stay the same.
 
Despite  the simplifying  assumptions, the  validity of  the adiabatic
approximation of  Blumenthal \etal (1986)  has been confirmed  down to
$10^{-2}r_{-2}$ in  a study of the  response of a dark  matter halo to
the  growth  of  an   exponential  disk  in  high-resolution  $N$-body
simulations (Jesseit  \etal 2002).  However, Wilson  (2003) and Gnedin
\etal  (2004) claim that  under more  general conditions  the standard
model  for  adiabatic   contraction  systematically  overpredicts  the
contraction in  the innermost regions,  while slightly underpredicting
the contraction at larger radii.  Therefore, we use the standard model
for adiabatic contraction  to provide an upper limit  on the effect of
adiabatic contraction.

 
\begin{figure*}[t] 
\begin{center}
\figurenum{5} 
\includegraphics[bb=  18 385 565 695, width=6.1in]{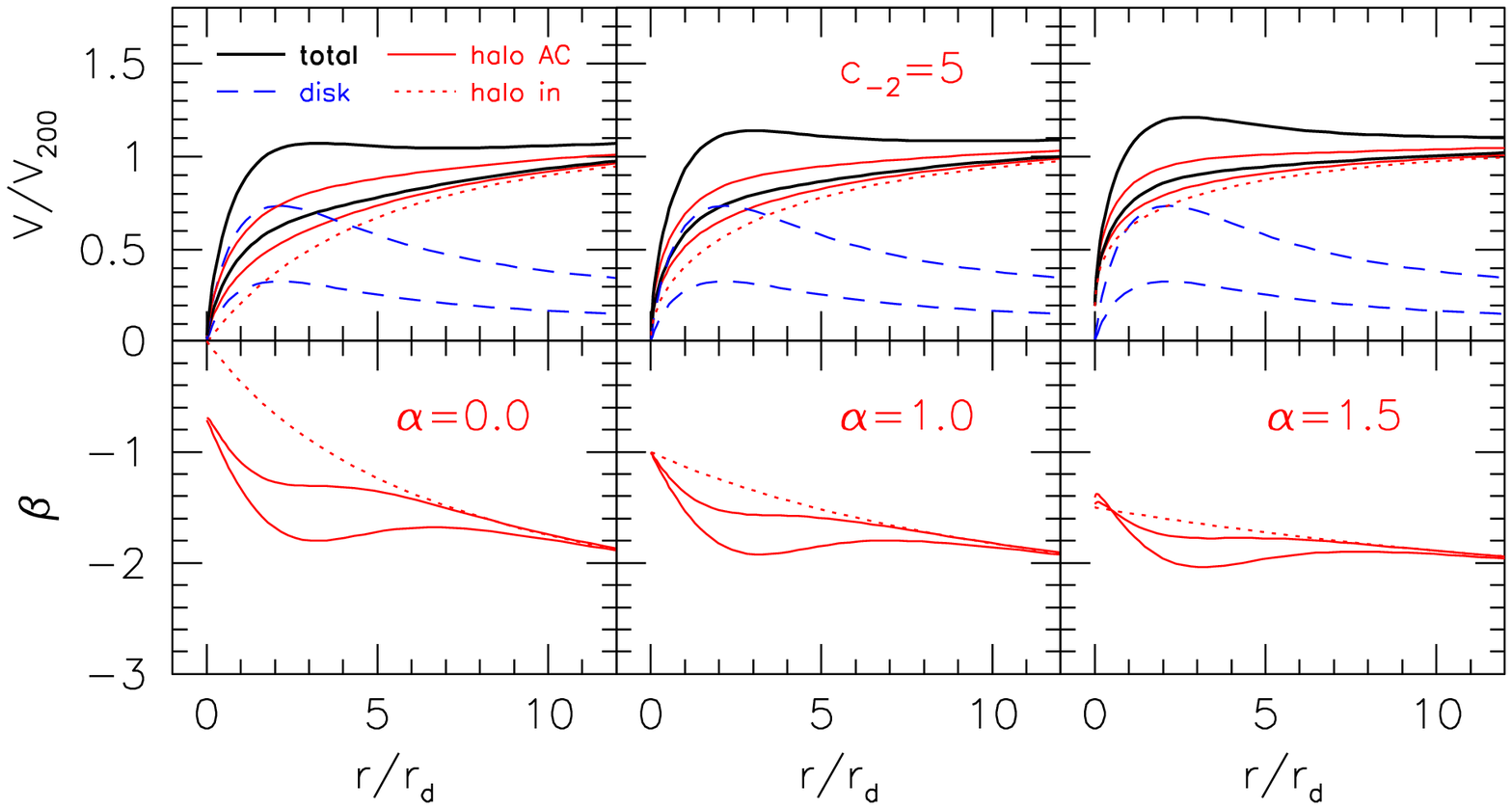}
\includegraphics[bb=  18 385 565 695, width=6.1in]{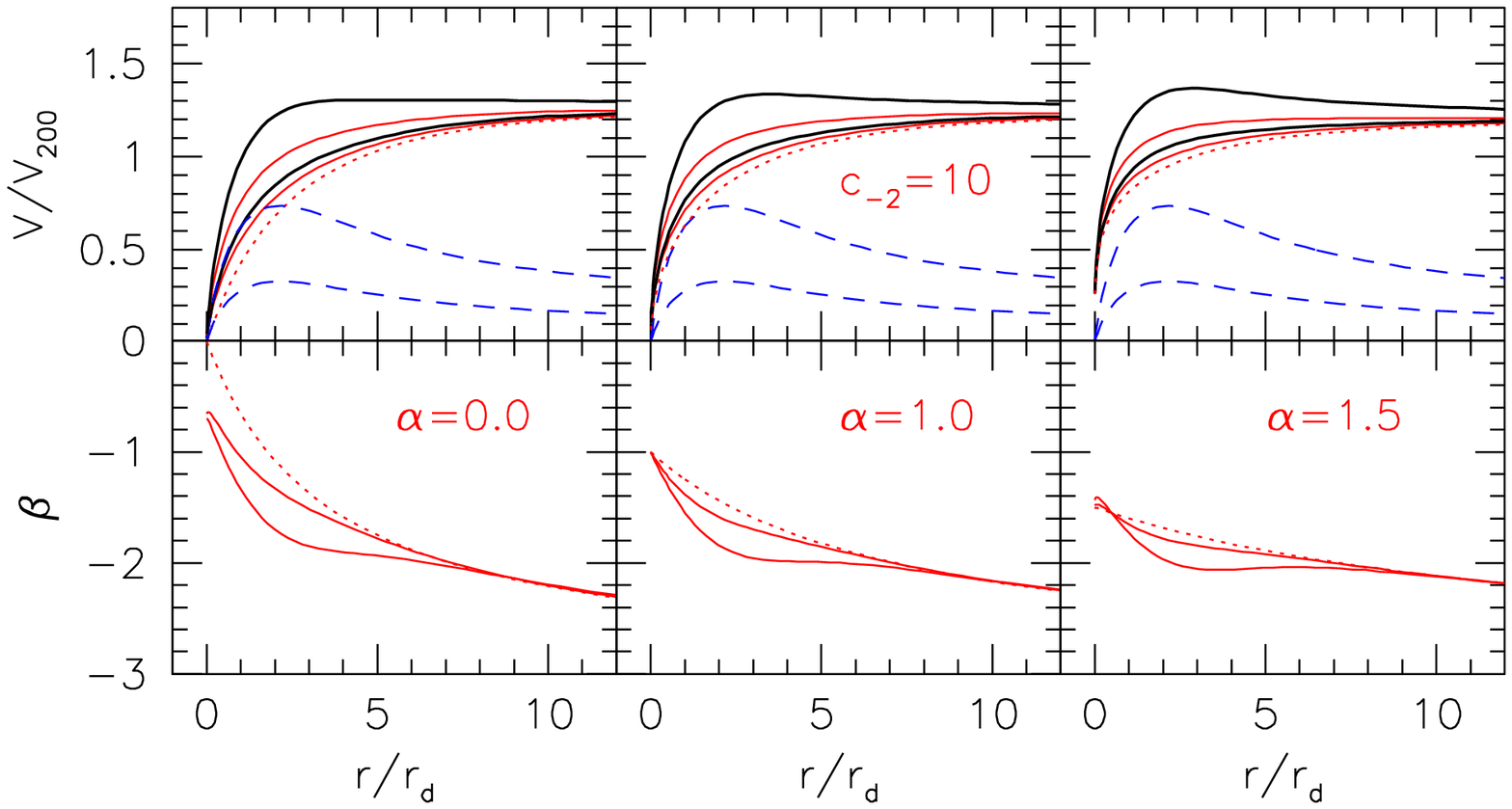}
\caption{Effect of adiabatic contraction on circular velocity and halo 
density slopes for  halos with $\alpha=$0, 1.0, and  1.5 ({\it left to
right}),  $c_{-2}=5$   (top)  and  10   (bottom),  for
exponential disks with $R_{\rm d} =$  2 kpc and $\Sigma_0 = $ 258 and
52 $\Msolpc2$.   For each  disk-halo system  we show the  initial halo
({\it red  dotted line}), final halo after  adiabatic contraction (AC;
{\it  red sold  line}),  disk  ({\it blue  dashed}),  and final  total
circular velocity ({\it black thick solid}).  }
\vspace{-0.2in}
\end{center}
\end{figure*}

\section{Model Degeneracies} 
Several  degeneracies exist  between the  model parameters,  which may
prevent  a  unique mass  decomposition.   These  can  be divided  into
``disk-halo'' and ``cusp-core'' degeneracies.
 
The  ``disk-halo''   degeneracy  occurs  between  \Yd   and  the  halo
parameters.   Equally  good fits,  in  a $\chisqr$-statistical  sense,
where $\chisqr$ is  the reduced-$\chisq$, can be obtained  with a wide
range  of \Yd  from zero  to a  maximum disk  (e.g., van  Albada \etal
1985).   To break  this  degeneracy,  we need  a  priori knowledge  of
$\Upsilon_{\rm  d}$, or  $\vdisk$/$\vtot$. If  the disk  thickness and
halo flattening  are ignored,  this fixes the  density profile  of the
dark  halo.   However,  in  practice,  with  errors  on  the  circular
velocities of  a few kilometers per second  finite spatial resolution,
and  limited extent of  the rotation  curve, degeneracies  between the
halo parameters themselves often  prevent a unique parameterization of
the halo  density profile (van den  Bosch \& Swaters  2001).  To break
this degeneracy, constraints  need to be placed on  \ccm2 and \v200 as
well.
 
We illustrate these degeneracies  with mock rotation curves.  Our mock
galaxies consist of an exponential disk specified by
$\mu_0^R=20~$mag~arcsec$^{-2}$, $\Rd=2~$kpc, and \YdR=1 and an
adiabatically   contracted   dark   halo   with   initial   parameters
$c_{-2}=10$, $V_{200}=100\ \kms$,  and $\alpha = 0$, 0.5,  1, and 1.5.
We then sample the rotation curve  in 3\arcsec\, bins up to 2$R_d$ and
in  15\arcsec\, bins up  to 8$R_d$  to simulate  \ha\, and  \hi\, data
respectively.   We then add  a random  Gaussian error  (with $\sigma=4
\kms$) and assign  a conservative measurement error of  4 \kms to each
data point.
 
We fit for \ccm2 and \v200 on  a grid of $\alpha$ and $\YdR$, with and
without adiabatic contraction. The results  of these fits are shown in
Fig.~6.  The disk-halo and  cusp-core degeneracies exist for all input
values of $\alpha$ and are  strongest for $\alpha=1$ halos.  Thus, for
these  model galaxies, {\it  without constraints  it is  impossible to
determine \YdR or  $\alpha$ based on the $\chisqr$  value alone}. When
fitting  without  adiabatic contraction,  a  wider  range  of \YdR  is
permitted, including maximum disks  for $\alpha=0$ halos.  We also see
that  the form of  the $c_{-2}-\alpha$  relation is  the same  for all
fits,  but  the  normalization   is  lower  for  fits  with  adiabatic
contraction and a higher input $\alpha$.
 
In  order  to  achieve  reliable  results out  of  the  mass  modeling
exercise, we must therefore consider independent constraints, which we
discuss below.
 
\begin{figure*}[t] 
\begin{center}
\vspace{0.5in}
\figurenum{6}
\includegraphics[bb=  20 410 570 720, width=7.0in]{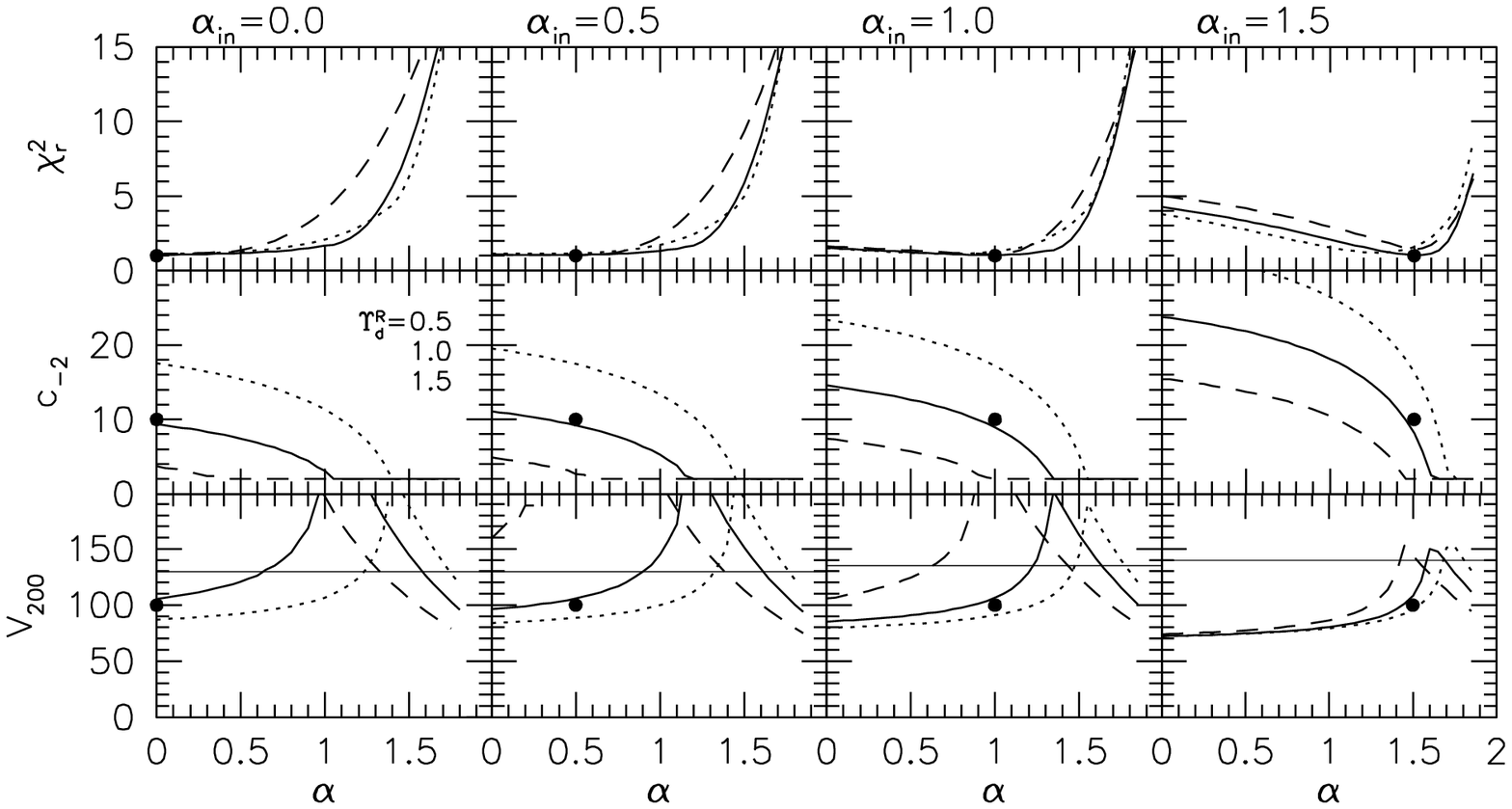}
\includegraphics[bb=  20 410 570 720, width=7.0in]{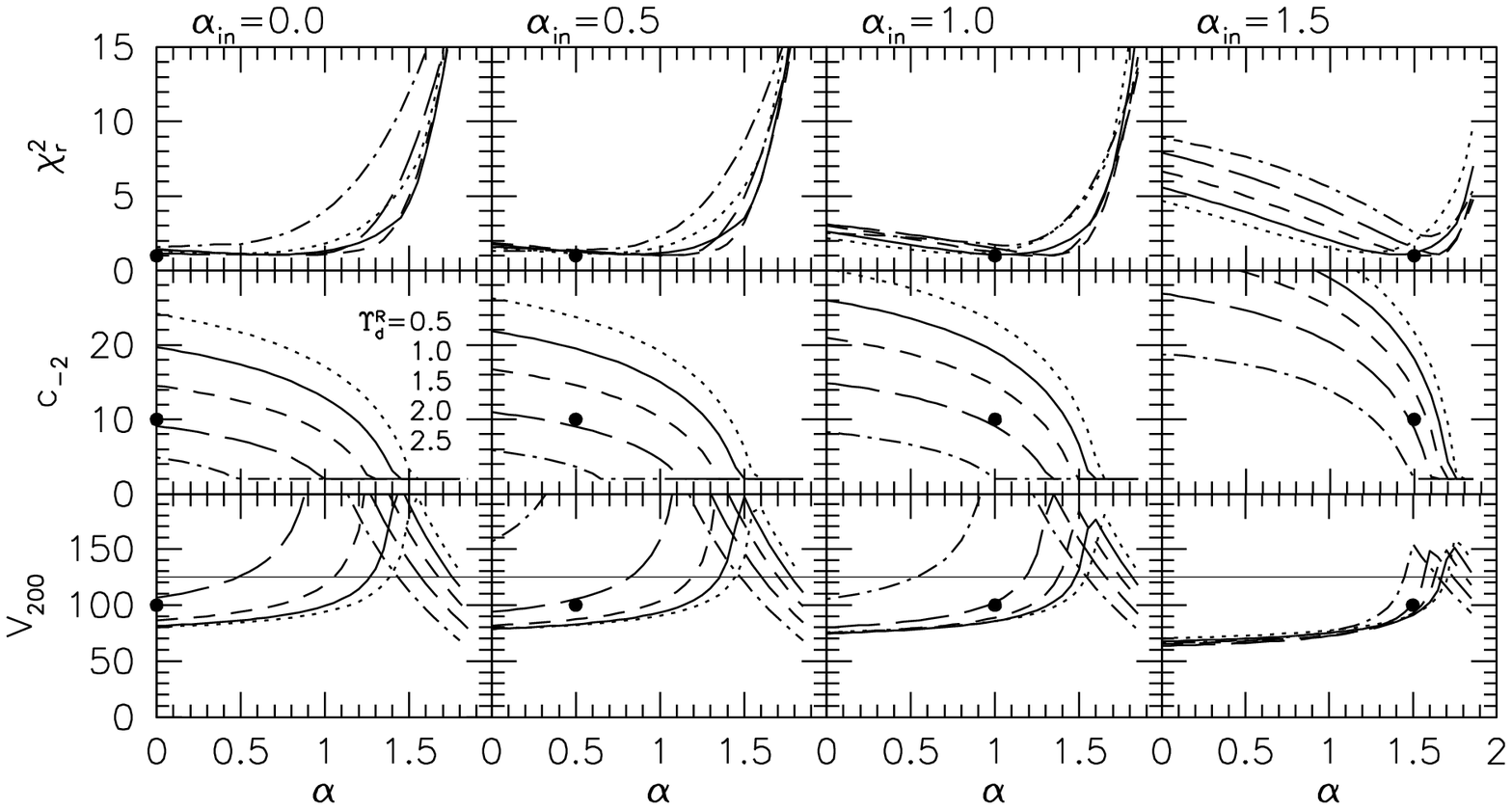}
\caption{Best-fitting halo parameters and $\chi^2_{\rm r}$ for mock rotation 
curves vs.  $\alpha$.  All  input models have adiabatically contracted
(AC) halos with $c_{-2}=10$, $V_{200}=100$, and exponential disks with
$\mu_0^R=20$ and  $\Upsilon_d^R=1.0$.  The  only difference is  in the
central density  slope, $\alpha$.  The  input models are  indicated by
the filled circles and have  $\chi^2_r=1.0$.  The top panels show fits
with AC,  while the  bottom panels show  fits without.   The different
lines   correspond  to  different   fitted  mass-to-light   ratios  as
indicated.   The horizontal  line indicates  the  maximum ``observed''
circular velocity.  Note that we have imposed $\c_{-2} \ge 2$.}
\vspace{0.5in}
\end{center}
\end{figure*}


\section{Constraints} 
\subsection{Stellar Population Synthesis Models} 
Stellar  population  synthesis  (SPS)  models  can be  used  to  place
constraints  on  $\Yd$.   The  combination  of  optical  and  infrared
photometry, with SPS models, yields \Yd values accurate to $\sim 40$\%
(Bell  \& de  Jong 2001).   The slope  of \Yd  versus color  is fairly
independent of the initial mass function (IMF) and star formation (SF)
history.  That slope is also smaller in the $K$-band  than  in the
$B$-band, although the zero point  of the color-\Yd relation is itself
very sensitive to the IMF.   The calibration of the \Yd-color relation
by Bell  \& de Jong (2001)  relies on the assumption  of some galaxies
being close to  maximal disks, and their values of  \Yd are thus upper
limits.

\begin{center}
\begin{deluxetable*}{ccccccccccccc}[t]  
\tablewidth{0pt}
\tabletypesize{\small} 
\tablecolumns{13} 
\tablenum{1} 
\tablecaption{Galaxy parameters}
\tablehead{ 
\colhead{Galaxy} &  
\colhead{$M_{B}$} & 
\colhead{$\;$Band$\;$}  &  
\colhead{$\mu_0^{\rm R,c}$} &  
\colhead{$R_{\rm d}$}  &  
\colhead{$D$} &  
\colhead{$V_{\odot}$} &  
\colhead{$i$} &  
\colhead{$1/\cos(i)$} & 
\colhead{$\;V_{\rm max}\;$} &  
\colhead{$\;R_{\rm H\alpha}\;$} & 
\colhead{$\;R_{\rm HI}\;$} & 
\colhead{References} \\ 
    & $\;\;$(mag)$\;\;$ &  & ($\rm mag\,arcsec^{-2}$) & $\;$(kpc)$\;$ &$\;$(Mpc)$\;$ &$\;$ ($\kms$$\;$) 
    & (deg) & & & & \\
(1) & (2) & (3) & (4) & (5) & (6) & (7) & (8) & (9) & (10) & (11) & 
(12) & (13) } 
\startdata 
NGC 3109..............& -16.35 & $B$ & 22.3 & 1.3  
& $1.36^{\c}$ & 403 & $75\pm5$&3.86  
& 67 & 2.1 & 5.0  
& 1, 2, 1, 3
\\ 
IC 2574..................& -16.77 & $R$ & 23.0 & 2.3  
& $4.0^{\b}$ & 57  & $75\pm7$&3.86 & 67 & $-$& 4.6  
& 4, 2, 4, 5 
\\ 
UGC 2259.............. & -17.03 & $r$ & 21.7 & 1.6  
& $10.3^{\h}$ & 583 & $41\pm3$& 1.33 & 90 & 1.4 & 4.8  
& 6, 7, 8, 7 
\\ 
NGC 5585.............. & -17.50 & $R$ & 21.1 & 1.9  
& $8.7^{\b}$ & 305 & $53\pm1$& 1.66 & 92  & 2.7 & 7.1  
& 9, 10, 9, 11  
\\ 
NGC 2403.............. & -19.50 & $r$ & 20.4 & 1.8  
& $3.22^{\c}$& 131 & $60\pm2$& 2.00 & 136 & 2.1 & 10.8   
& 12, 7, 8, 13 
\\               
NGC 3198.............. & -19.90 & $r$ & 21.0 & 3.7  
& $13.8^{\c}$& 663 & $72\pm2$& 3.24 & 157& 1.8 & 12.0  
& 12, 14, 8, 13 
\enddata 
\tabletypesize{\footnotesize}
\tablecomments{\footnotesize
Col. (1): Galaxy name.  
Col. (2): Absolute $B$ magnitude (Blais-Ouellette 2000). 
Col. (3): Photometry band; where necessary we convert to Cousins $R$  
assuming $B-R=0.9$ or $r-R=0.35$ (Jorgensen 1994).  
Col. (4): Central surface brightness in $R$ band from a fit to the
surface brightness profile with a marked disk, corrected for
inclination (Col.~[8]) and Galactic extinction (Schlegel \etal 1998).  
Col. (5): Scale length of the disk, from a fit to the $R$-band surface brightness profile with a marked disk. 
Col. (6): Adopted distance, with distance indicators  
``c'' for Cepheid, ``b'' for brightest stars, and ``h'' for Hubble distance ($h=0.7$) corrected for Virgo-centric flow (150 km/s). 
Col. (7): Systemic velocity; see \hi\, reference. 
Col. (8): Inclination; see \hi\, reference. 
Col. (9): Conversion factor we use to correct our $\Upsilon_{\rm d}$ for inclination, assuming a thin disk with no extinction. 
Col. (10): Maximum rotation velocity; see \hi\, reference. 
Col. (11): Radius out to which we use the \ha\, rotation curve  
(in $R_{\rm d}$). 
Col. (12): Radius out to which we use the \hi\, rotation curve  
(in $R_{\rm d}$). 
Col. (13): References, in the following sequence: \hi, \ha,
photometry, D.}
\\
\tablerefs{\footnotesize
(1): Jobin \& Carignan (1990); 
(2): Blais-Ouellette et al. (2001);  
(3): Musella et al. (1997); 
(4): Martimbeau et al. (1994); 
(5): Karachentsev et al. (2002); 
(6): Carignan et al.(1988); 
(7): Blais-Ouellette et al. (2004);  
(8): Kent (1987);   
(9): C\^{o}t\'{e} et al. (1991). 
(10): Blais-Ouellette et al. (1999);  
(11): Drozdovsky \etal (2000);  
(12): Begeman (1987);  
(13): Freedman \etal (2001);  
(14): Corradi et al. (1991);  
}  
\end{deluxetable*}
\end{center} 

\vspace{-0.8cm} 
\subsection{Evidence for Sub-Maximal Disks} 
A conventional hypothesis  to determine an upper limit  to \Yd is that
disks should be maximal\footnote{We  adopt the definition of a maximal
disk  as one  that  supplies $85\pm  10\%$  of the  total velocity  at
$2.2\Rd$ (Sackett  1997).}  (Carignan \&  Freeman 1985; van  Albada \&
Sancisi  1986).  This  approach works  well in  practice for  most HSB
galaxies,  but  dark matter  is  still  needed  to explain  the  outer
rotation curves of HSB and LSB  galaxies at almost all radii. The fact
that maximum disks can match  inner rotation curves of HSB galaxies is
more telling about the degeneracies in mass modeling than the validity
of the  hypothesis itself (Broeils  \& Courteau 1997; Courteau  \& Rix
1999).  Furthermore, \ha\ rotation curves alone can often be fitted by
pure disk  or pure  halo models and  thus lack any  constraining power
without  the addition  of an  extended \hi\  rotation  curve (Buchhorn
1992; Broeils \& Courteau 1997).
 
 By contrast  to the  maximal disk hypothesis,  a variety  of methods,
which  are described  below, suggest  that  on average  HSB disks  are
sub-maximal with
\begin{equation} 
\left(V_{\rm disk}/V_{\rm tot}\right)_{2.2} \simeq 0.6.
\end{equation} 
Note that a  galaxy with a sub-maximal disk at $2.2  \Rd$ can still be
baryon  dominated  at  $2.2  \Rd$  if there  is  a  significant  bulge
component.

\vspace{-0.2cm} 
\subsubsection{TFR Residuals} 
Courteau \& Rix (1999) have suggested that sub-maximal disks should be
invoked  to  explain  the   surface  brightness  independence  of  the
Tully-Fisher  relation (TFR); they  find that,  {\it on  average}, HSB
galaxies have  $\left(V_{\rm{disk}}/V_{\rm{obs}}\right)_{2.2} \lta 0.6
\pm 0.1$ (see also Courteau  \etal 2005).  Their argument only depends
on the assumptions that the scatter in the TFR and the size-luminosity
relation (SLR)  is dominated  by a dependence  in $\Rd$ and  that dark
halos respond adiabatically to the formation of the disk.

\subsubsection{Velocity Dispersion Measurements} 
The  peak circular  velocity of  an isolated  exponential disk  can be
related to the vertical velocity dispersion\footnote{The correction to
the velocity  dispersion of a disk  embedded in a dark  matter halo is
usually negligible (Bottema 1993).} and the intrinsic thickness of the
disk, $z_0$, via (Bottema 1993)
\begin{equation} 
V_{\rm{disk}}^{\rm    max}    =    0.88   \,    {\left\langle    V_z^2
\right\rangle}^{1/2}_{R=0} \,\sqrt{\frac{R_{\rm{d}}}{z_0}}.
\end{equation} 
The factor  0.88 applies to  a disk of  zero thickness; for  a thicker
disk the peak velocity will be lower.
 
In practice, these measurements  are difficult since the scale height,
$z_0$,  and  scale length,  $\Rd$,  of  the  disk cannot  be  measured
simultaneously and  the vertical  velocity dispersions are  easiest to
measure in  face-on galaxies although  the disk kinematics is  hard to
determine.   The prospects  for  this method  are  improving with  the
ability  to reliably  determine the  inclinations  for low-inclination
galaxies using integral field spectroscopy (Verheijen \etal 2004).
 
Bottema (1993) found, for  stellar kinematic measurements in 12 spiral
galaxies (with $V_{\rm obs}^{\rm{max}} > 100 \,\rm{km}\,\rm{s}^{-1}$),
that more  massive spirals have  larger velocity dispersions  with the
correlation     ${\left\langle     V_z^2    \right\rangle}^{1/2}_{R=0}
={\left\langle V_R^2 \right\rangle}^{1/2}_{R=R_d} = (0.30 \pm 0.06) \,
V_{\rm{obs}}^{\rm{max}}.$  Substituting  this  into equation~(11)  and
taking  the  intrinsic  disk  scale  ratio  $R_{\rm  d}/z_0=4.2\pm1.5$
(Kregel \etal~2002) yields
\begin{equation} 
V_{\rm{disk}}^{\rm{max}}/V_{\rm{obs}}^{\rm{max}} = 0.5 \pm 0.2.
\end{equation} 
By comparison,  Bottema (1993)  obtained a mean  value of  0.63, using
$R_d/z_0=6$.

\subsubsection{Gravitational Lensing} 
In some rare cases in which  a quasar is lensed by a foreground galaxy
and gravitational lensing can be  used to place an extra constraint on
the mass  profile, the dynamical analysis  strongly favors sub-maximal
disks (Maller et al. 2000; Trott \& Webster 2002).

\subsubsection{Bars and Spiral Structure} 
It is  generally thought that  dynamical friction between the  bar and
halo will  slow down  the pattern  speed of the  bar; thus,  fast bars
imply  maximal  disks  (Weinberg  1985; Hernquist  \&  Weinberg  1992;
Debattista \&  Sellwood 2000).  However, other authors  claim that the
efficiency   of  bar   slow-down  by   dynamical  friction   has  been
overestimated  (Valenzuela \& Klypin  2003) and  that the  bar pattern
speed  is  not  a  reliable  indicator of  disk-to-dark  matter  ratio
(Athanassoula  2003).   Current  observational  data  favor  fast  bar
pattern  speeds; however,  only a  handful of  galaxies  have reliable
measurements   (e.g.,   Debattista  \&   Williams   2004),  and   most
observations are of SB0  spiral galaxies, whose large bulge components
and red colors are consistent with being baryon dominated.

In terms of late-type spiral galaxies, Weiner \etal (2001) modeled the
strong shocks  and non-circular motions  in the observed gas  flow and
find that a high $\Yd$,  corresponding to 80\%-95\% of $\vdisk$, and a
fast-rotating bar are highly favored.   On the other hand, modeling of
the spiral arm structure of a few grand-design galaxies by Kranz \etal
(2003) yields a  wide range of $\vdisk$/$\vtot$,  from closely maximal
to 0.6. It should be noted that both these methods are model dependent
and that  while the  basic dynamics of  bars and spiral  structure are
understood, there are issues that remain to be resolved.

Courteau  \etal  (2003) showed  that  barred  and non-barred  galaxies
belong to  the same TFR.   Thus, if the  argument by CR99  is correct,
barred  galaxies would,  on average,  harbor sub-maximal  disks.  Note
that  this is  consistent with  the  above observations,  if there  is
significant  scatter  in   $\vdisk$/$\vtot$,  or  if  $\vdisk$/$\vtot$
increases with surface brightness.

\subsection{Constraints on $V_{200}$} 
As  shown in  Figure~6, the  fitted  \v200 often  exceeds the  maximum
rotation velocity of the galaxy; by restricting $\v200 \le \vmax$, the
parameter space  is reduced. For  observed galaxies we  expect $V_{\rm
max}^{\rm  obs} \ge  V_{\rm  max}$ provided  that  the rotation  curve
flattens out  or declines at large  radii, as is  typical for extended
rotation curves of spiral galaxies (Casertano \& van Gorkom 1991).
 
The combined analysis of  galaxy-galaxy lensing from the Sloan Digital
Sky Survey (SDSS) and the TFR  led Seljak (2002) to postulate that the
rotation  velocity  of  early-  and late-type  $\sim  L^{*}$  galaxies
decreases  significantly from  its peak  value at  the  optical radius
($3.2\Rd$) to the virial radius $R_{200}$, with
\begin{equation} 
V_{\rm{obs}}^{\rm{max}}/\, V(R_{200} ) \simeq 1.8,
\end{equation} 
and a $2  \sigma$ lower limit of 1.4.  This  implies that the rotation
curve  declines  at large  radii  and  is  thus inconsistent  with  an
isothermal profile ($\rho\propto  r^{-2}$), unless the velocity excess
over $V(R_{200})$ is  due entirely to the stellar  component, which is
unlikely for late-type galaxies.  This result is consistent with Prada
\etal (2003), who find $\rho \propto r^{-3}$ at large radii.
 
Combining  this constraint  with  the above  evidence for  sub-maximal
disks, we  get the interesting result  that the total  velocity at the
virial radius is approximately equal to the peak velocity of the disk,
\begin{equation} 
V_{\rm{disk}}^{\rm{max}} \simeq V(R_{200}) \simeq V_{200}.
\end{equation}

\subsection{Halo Concentration Parameter} 
Cosmological simulations  suggest a  correlation between $\ccm2  $ and
$\Vvir$, such that more massive halos have lower concentrations (e.g.,
Bullock \etal  2001; Eke \etal  2001; Wechsler \etal 2002;  Zhao \etal
2003).  This  is because halos  with smaller masses  collapse earlier,
when the universe has a higher mean density. In the standard $\Lambda$
CDM cosmology ($\sigma_8=1.0, \OmegaM=0.3$) for $\v200 =100\,\kms$ the
mean $c_{-2}\simeq12$, but with a significant scatter of $\Delta\,\log
c_{-2}=0.14$  (Wechsler  etal  2002).    The  mean  \ccm2  is  roughly
proportional  to $\sigma_8$ and  is also  weakly dependent  on \OmegaM
such  that lower  \OmegaM gives  lower  $\ccm2 $.   Given the  current
observational  uncertainties in  these parameters,  the  mean $c_{-2}$
could easily be lowered by 25\%.  In the model of Bullock \etal (2001)
for $40 <  V_{200} < 160$, the $2\sigma$ range  in concentration is $6
\lta c_{-2} \lta 30$.


\section{Data} 
We now apply these constraints to  a test sample of disk galaxies from
Blais-Ouellette (2000).  This  sample is one of the  few with rotation
curves  derived  from  both  two-dimensional \ha\  and  \hi\  velocity
fields.  The  \ha\ data are needed  to probe the inner  rising part of
the  rotation   curve,  where  resolution  effects   often  limit  the
reliablity  of \hi\  data.  However,  \hi\ rotation  curves  are still
needed  to probe  the  outer  part of  the  rotation curve  (typically
extending to twice the optical  radius) and are essential to constrain
the halo mass.
The  full Blais-Ouellette  sample contains  10 galaxies,  with  a wide
range  of  luminosities and  surface  brightnesses.   We restrict  our
analysis to six of these galaxies that do not have significant bulges.
Accurately decomposing  the disk and bulge  components and determining
radial  \Yd gradients requires  near-IR imaging  (e.g.  $K$  band) and
optical  color  profiles.   We  will  return  to  these  issues  in  a
forthcoming paper.
Table~1  gives the  optical  and kinematic  parameters  of the  sample
galaxies,  as well  as references  for the  data sources.   While this
sample is by no means complete, it provides a representative selection
of LSB and HSB ``bulge-less''  disk galaxies against which we can test
our constraints.

\subsection{Rotation curve errors} 
Ideally, the  rotation curve errors would be  normally distributed and
indicate the uncertainty  in the circular velocity at  a given radius.
In practice, the errors are often defined in some ad~hoc fashion, such
as assigning a constant value to each velocity bin, and cannot be used
to place confidence levels on fitted model parameters.  We assume that
the observed rotation  velocity is equal to the  circular velocity and
therefore  that non-circular motions  (e.g., bars,  streaming motions,
pressure support) are not significant.
 
Blais-Ouellette (2000)  computes errors  on the \ha\,  rotation curves
using  $\sigma/\sqrt{N}$ in  each ring  of a  tilted ring  fit  to the
velocity field.  By contrast, the  errors on the \hi\, rotation curves
in our sample  are computed using the velocity  difference between the
approaching and  receding sides of  the galaxy.  To be  consistent, we
re-compute  the errors  on the  \ha\,  rotation curves  by taking  the
maximum of  $\sigma/\sqrt{N}$ and the velocity  difference between the
two sides  weighted by  the number  of points on  each side.   We also
impose  a  minimum error  of  2\kms (to  be  consistent  with the  \hi
observations), although the actual uncertainty is probably larger.  We
investigate the effect of a larger minimum error value in \S 8.

\begin{figure*}[t] 
\begin{center}
\vspace{0.5in}
\includegraphics[bb=  20 410 570 720, width=7.1in]{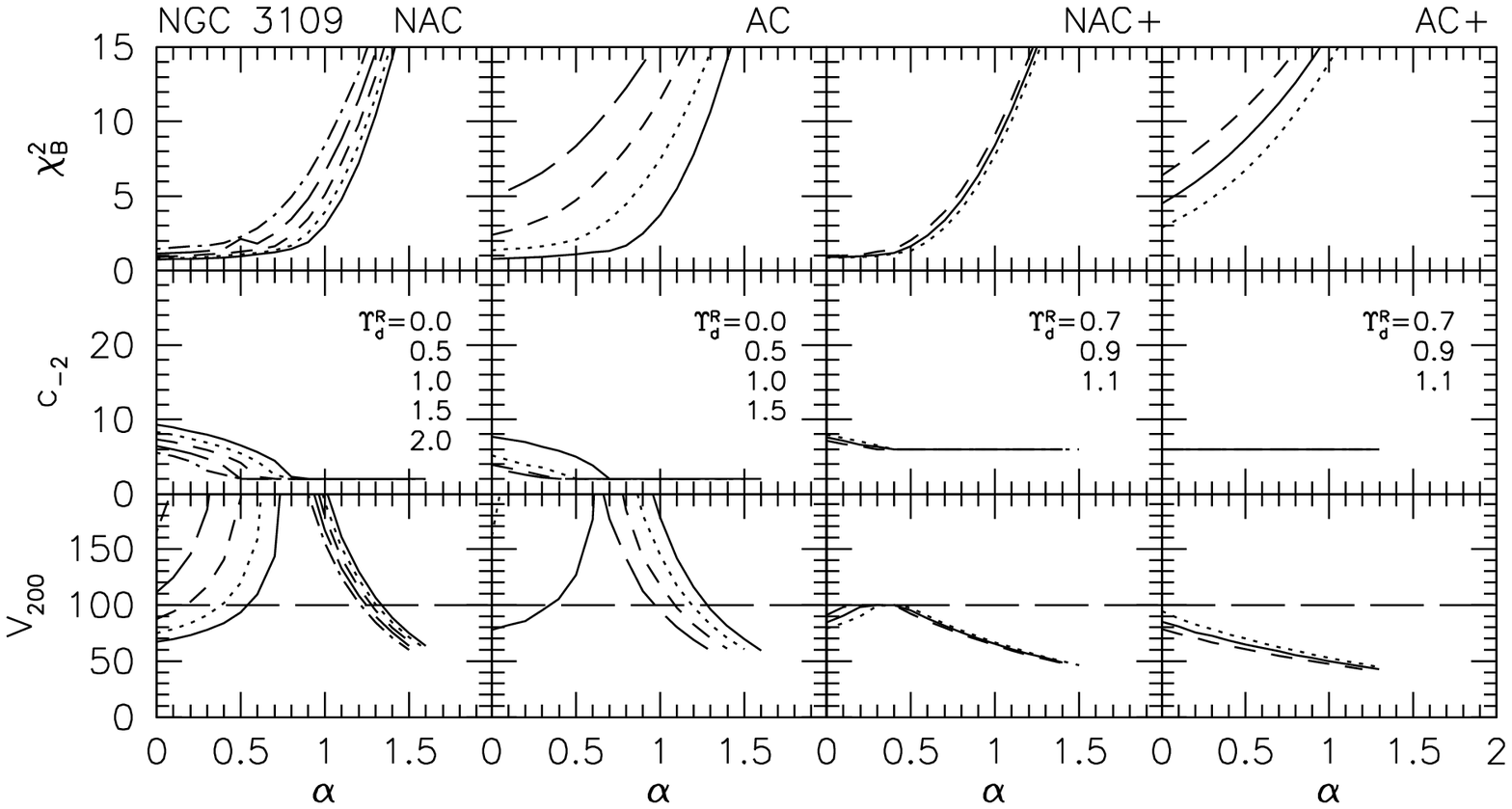}
\includegraphics[bb=  20 410 570 720, width=7.1in]{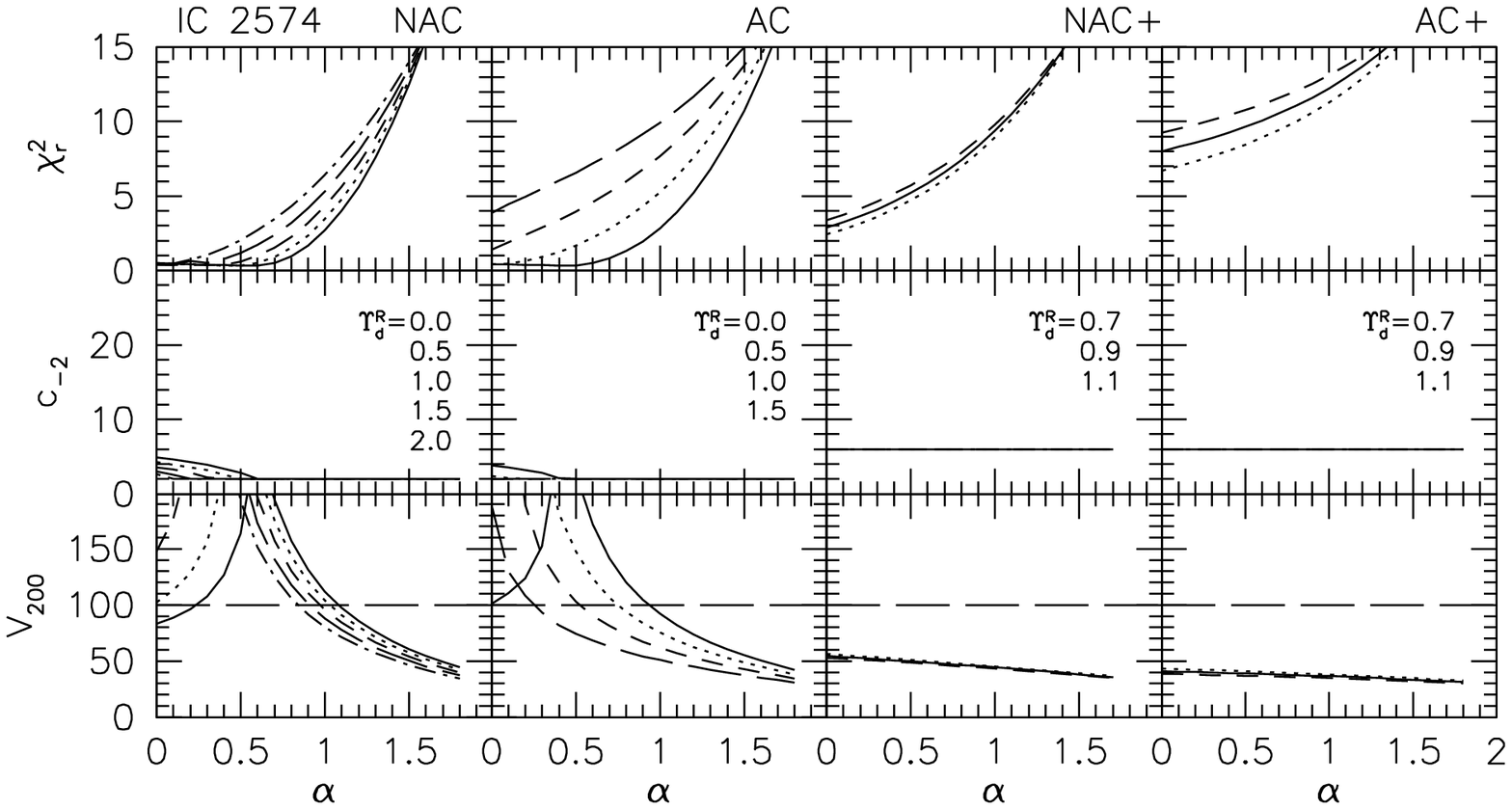}
\figurenum{7}
\caption{Best-fitting halo parameters and \chisqr vs. $\alpha$ for 
a range  of \YdR (as specified  in the figures).  We  show fits with
(AC) and without (NAC) adiabatic contraction and with (+) and without
constraints on \YdR, $ \ccm2  $, and $V_{200}$.  The dashed horizontal
line indicates the constraint  on $V_{200}$, from the maximum observed
velocity; for NGC 2403 and NGC 3198 we also show $V_{\rm max}/1.4$ and
$V_{\rm max}/1.8$.  For  NGC 3109 and IC 2574  the rotation curves are
still rising at the last measured point; here we set $V_{200} < 100$.}
\vspace{0.5in}
\end{center}
\end{figure*} 

\begin{figure*}[t]
\begin{center}
\vspace{0.5in}
\includegraphics[bb=  20 410 570 720, width=7.1in]{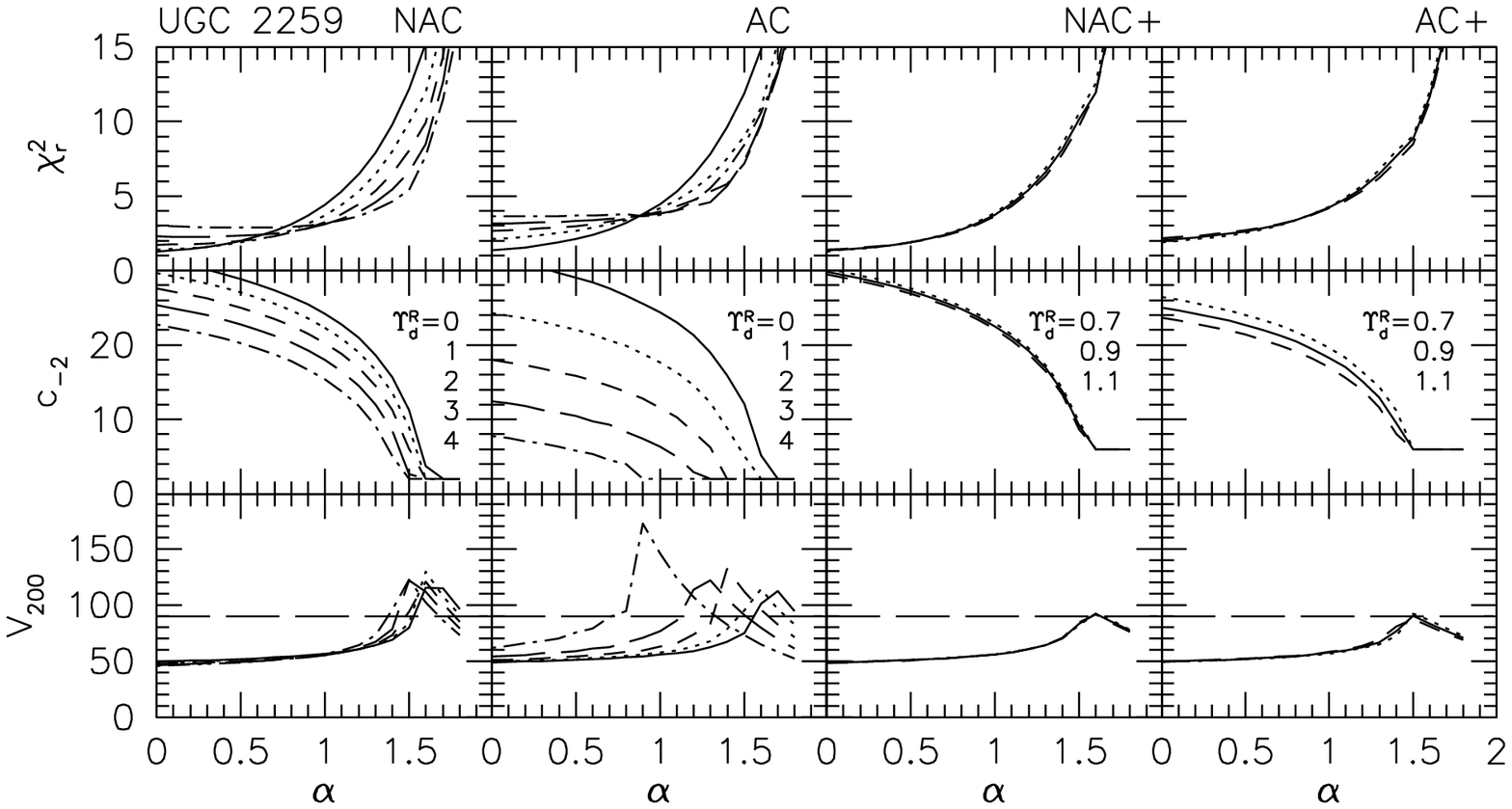}
\includegraphics[bb=  20 410 570 720, width=7.1in]{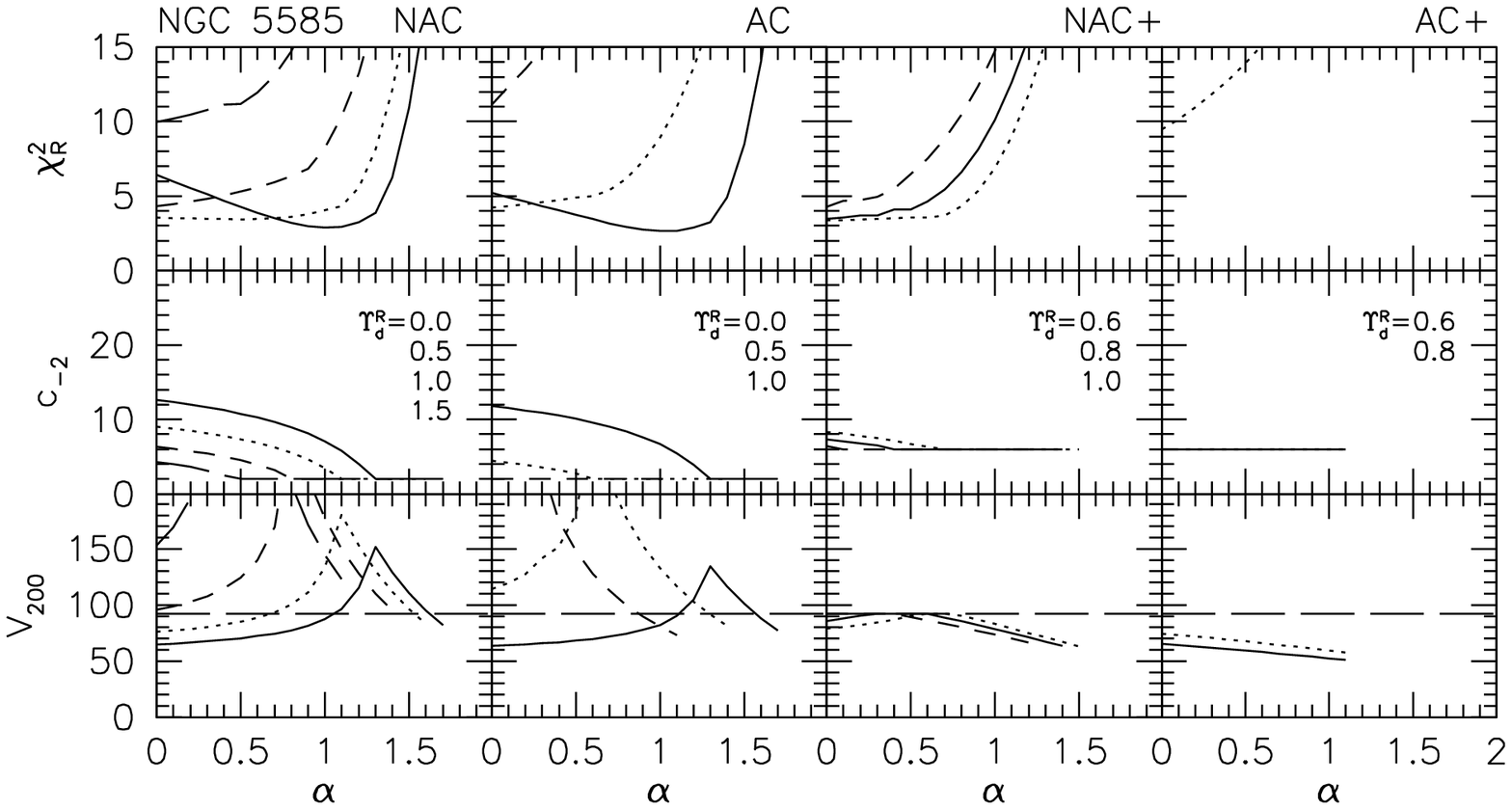}
\figurenum{7}
\caption{Continued.}
\vspace{0.5in}
\end{center}
\end{figure*} 

\begin{figure*}
\begin{center}
\vspace{-0.1in}
\includegraphics[bb=  20 410 570 720, width=7.1in]{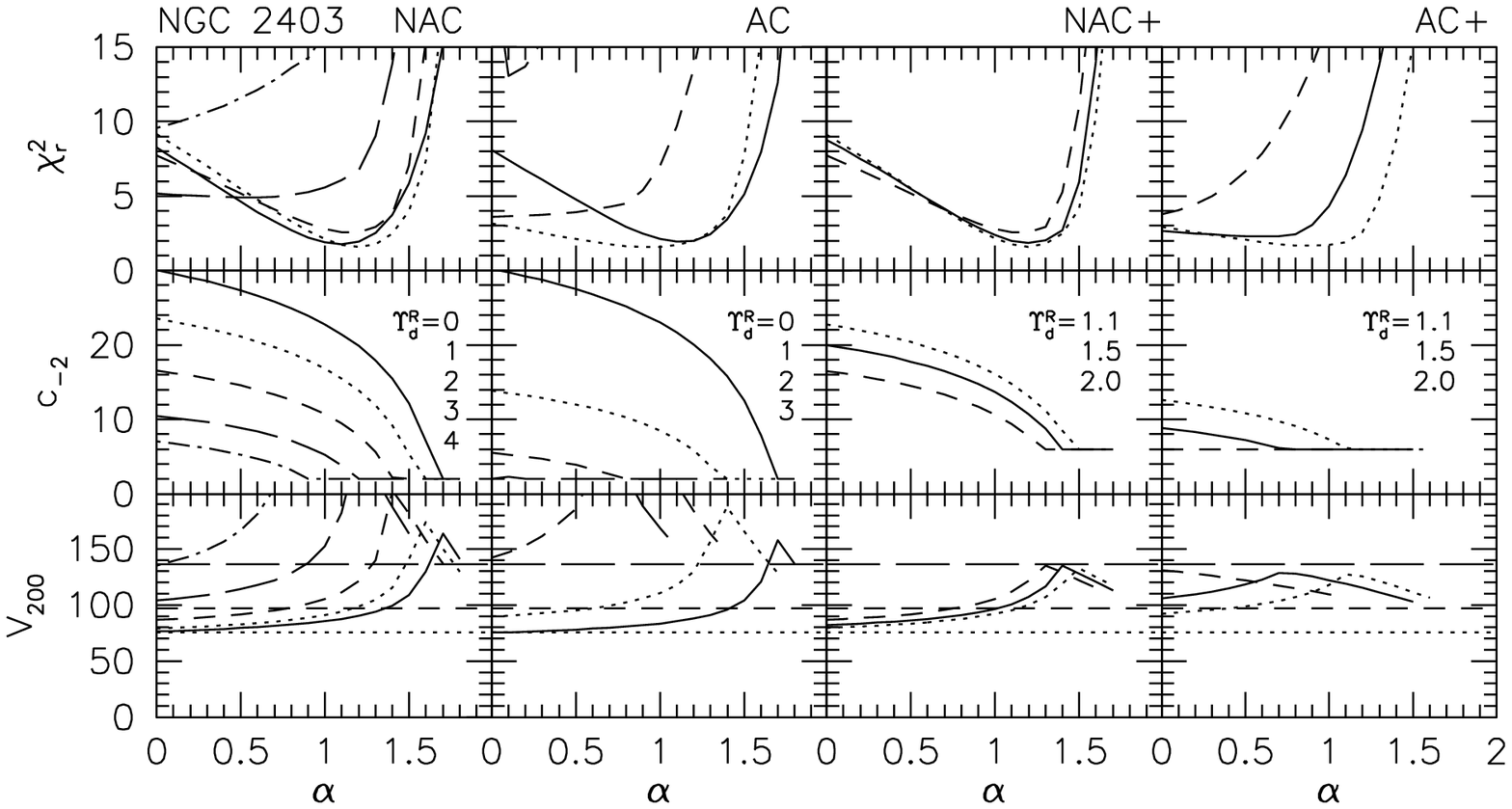}
\includegraphics[bb=  20 410 570 720, width=7.1in]{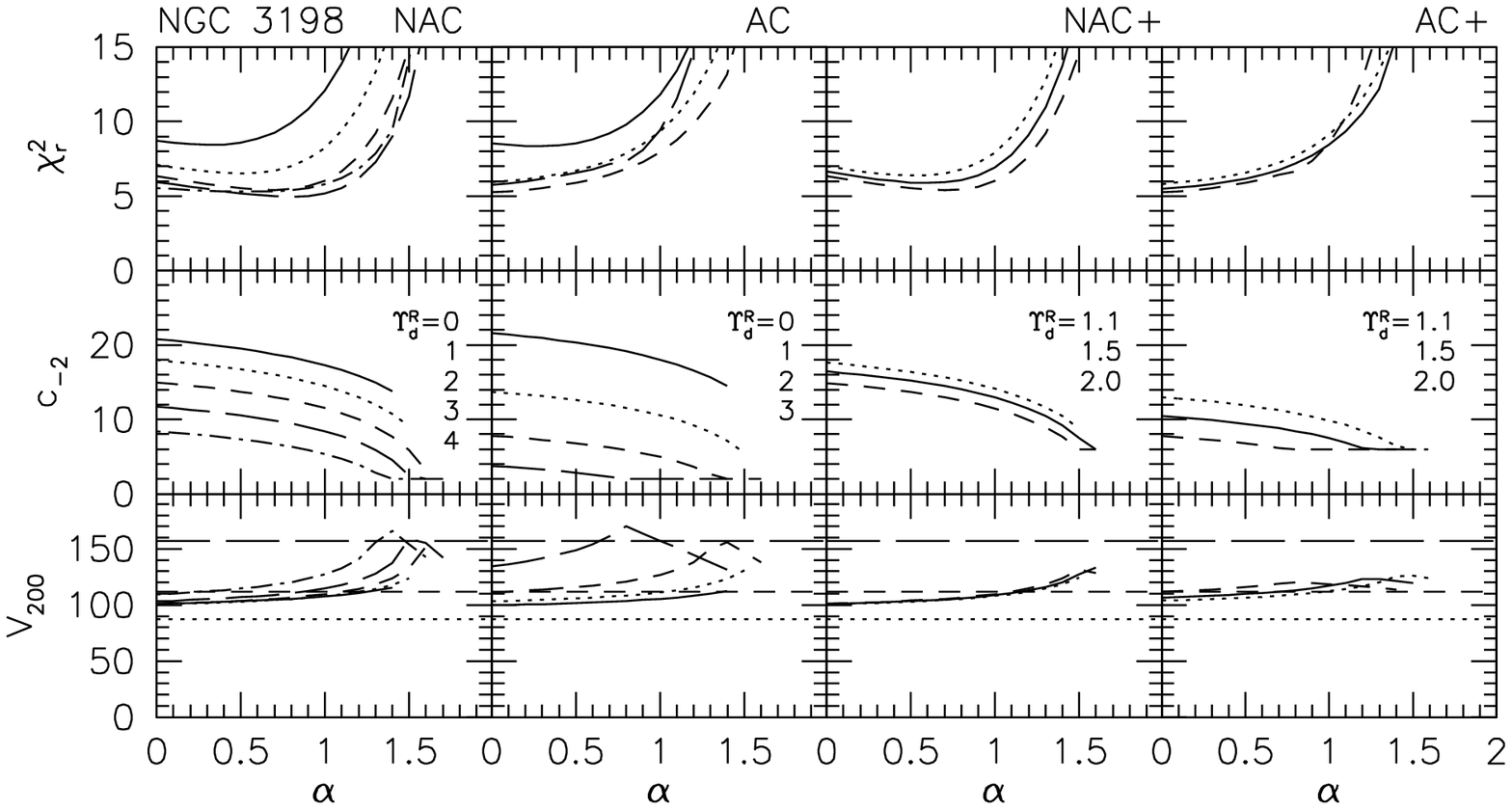}
\figurenum{7}
\caption{Continued.}
\vspace{-0.2in}  
\end{center}
\end{figure*}

\begin{figure*} 
\begin{center}
\includegraphics[bb=  25 360 580 715, width=7.0in]{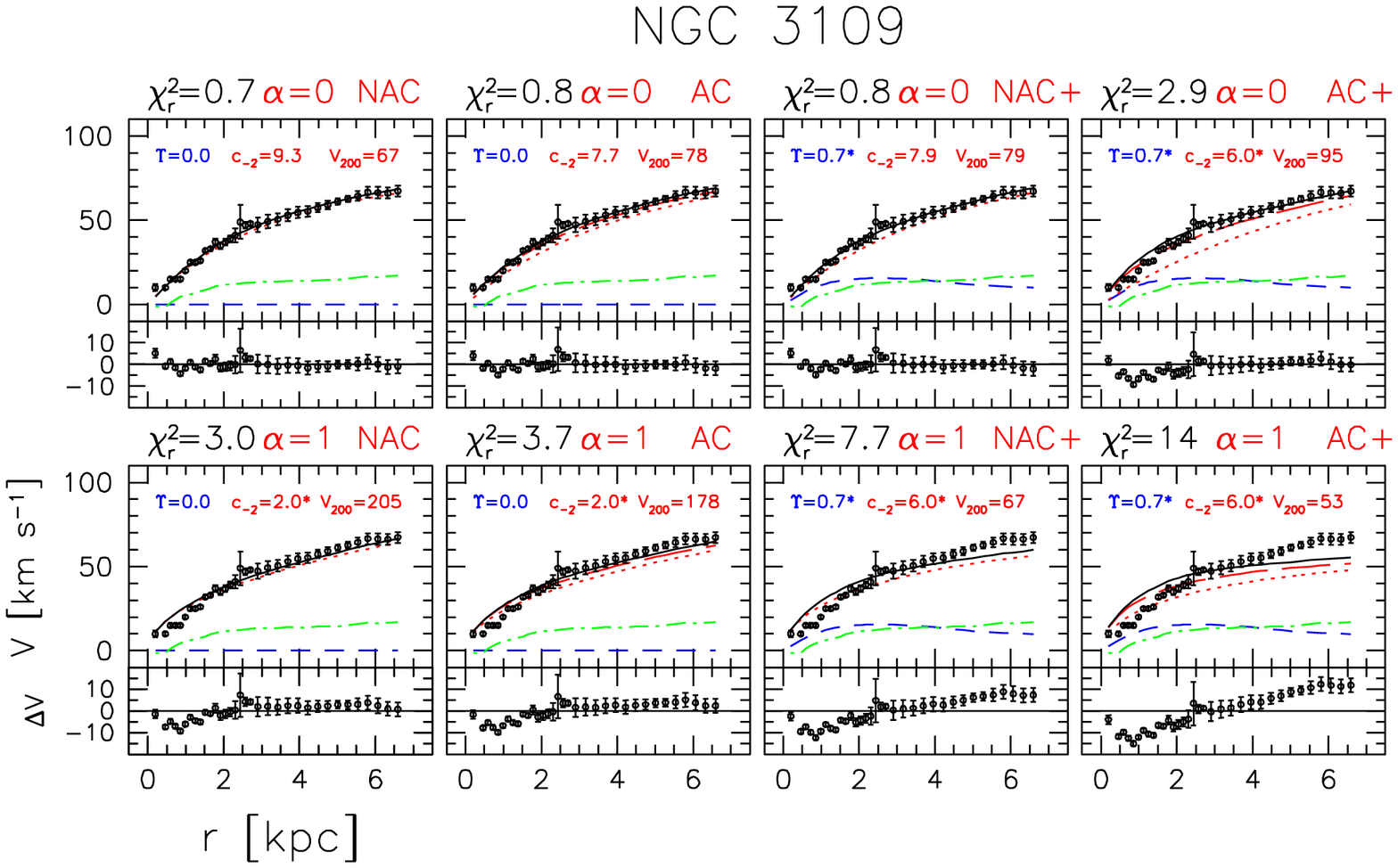}
\includegraphics[bb=  25 360 580 715, width=7.0in]{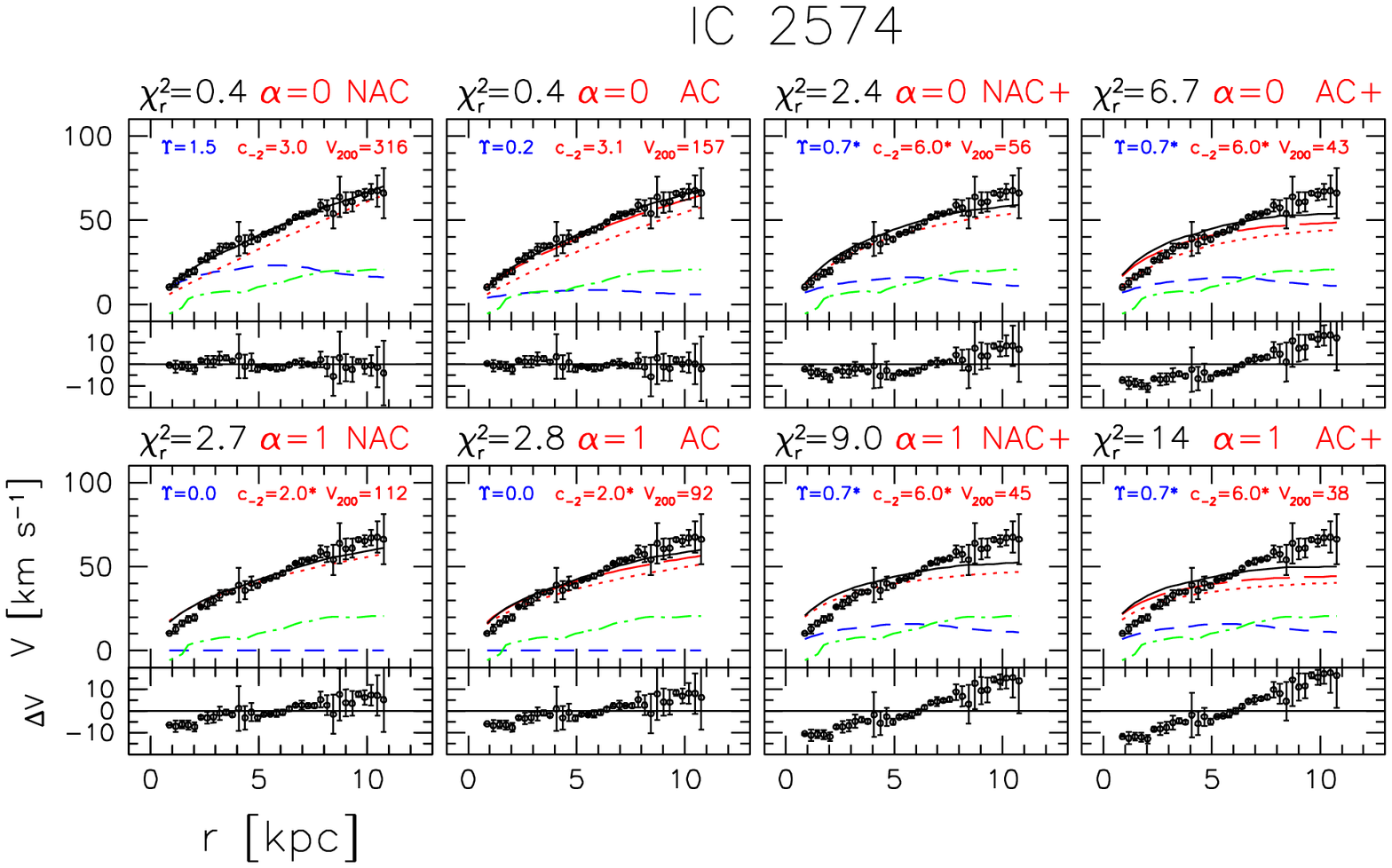}
\figurenum{8} 
\caption{ Sample rotation curve decompositions for all six galaxies.  
Shown are best fits (minimum \chisq) for $\alpha=0$ and $\alpha=1$, 
with (AC) and without (NAC) adiabatic contraction and with (+) and  
without constraints on $\YdR$, $c_{-2}$, and $v_{200}$. 
The total model rotation curves are given by the solid black lines; 
each model consists of three components: stellar disk ({\it blue
  dashed line}), gaseous disk ({\it green dot-dashed line}),  
and halo before/without AC ({\it red dotted line}) and after AC ({\it
  red long-dashed line}).}
\end{center}
\end{figure*}

\begin{figure*}
\begin{center} 
\includegraphics[bb=  25 360 580 715, width=7.0in]{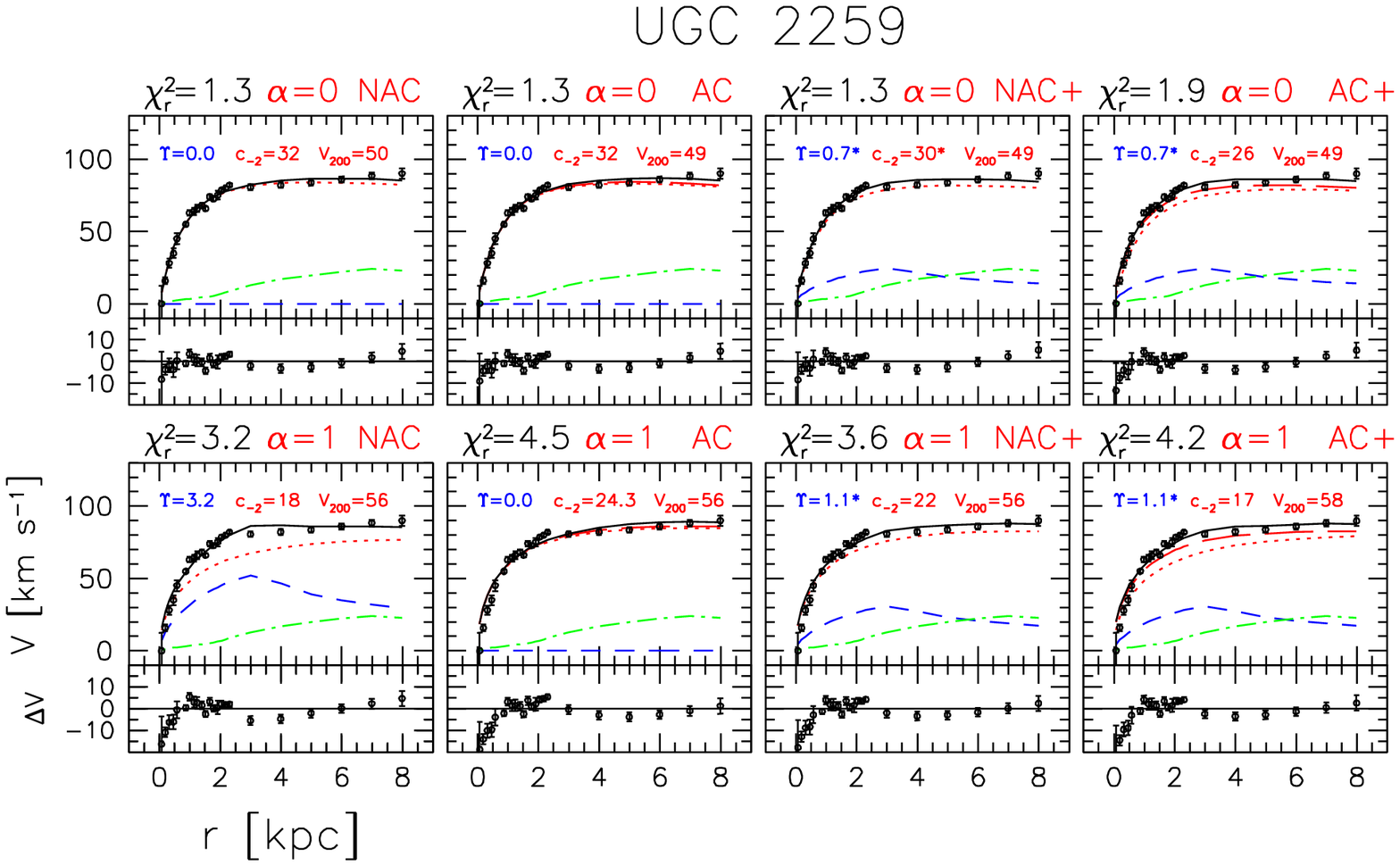}
\includegraphics[bb=  25 360 580 715, width=7.0in]{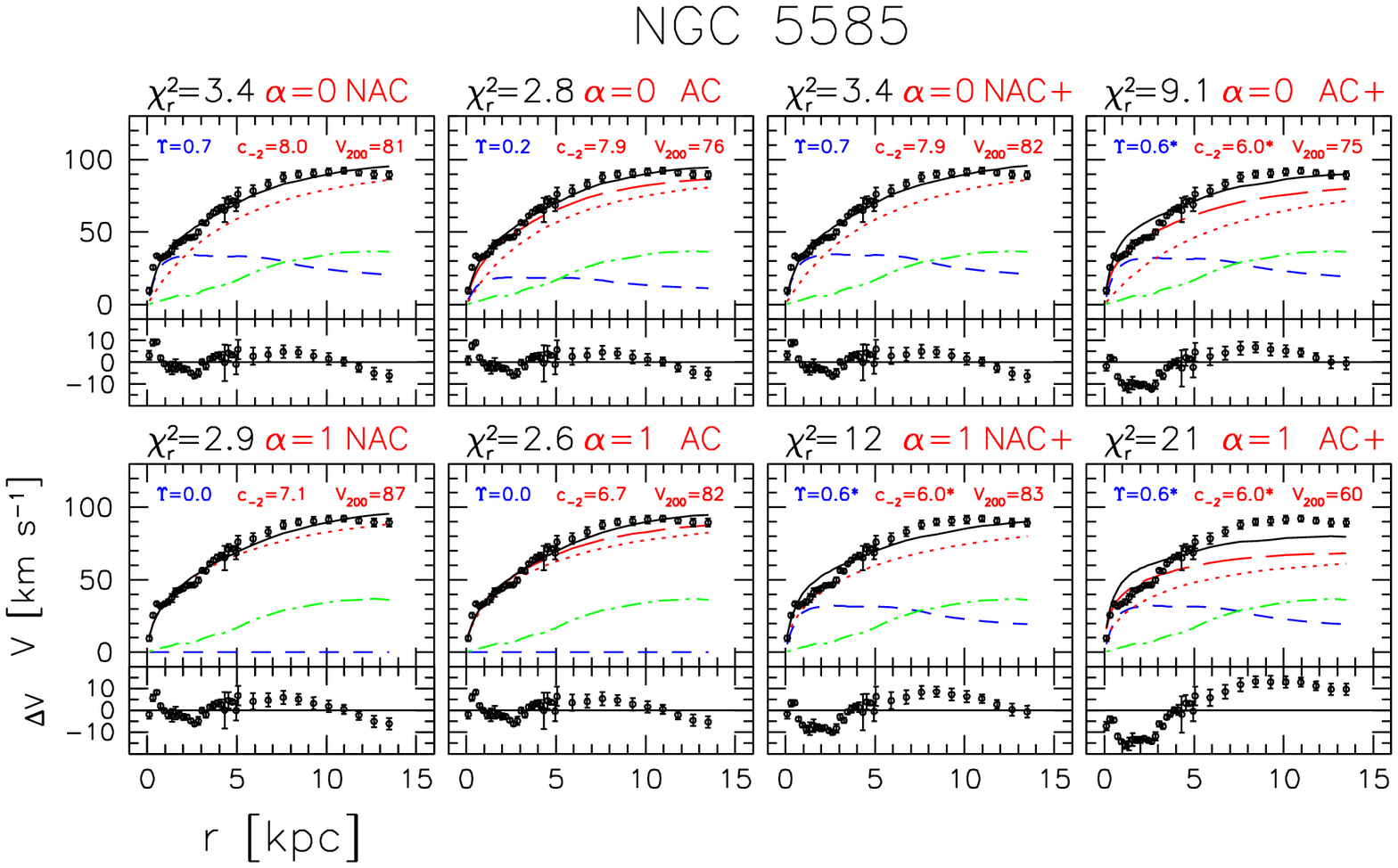}
\figurenum{8} 
\caption{Continued.}  
\end{center}
\end{figure*} 

\begin{figure*}[t]
\begin{center}
\includegraphics[bb=  25 360 580 715, width=7.0in]{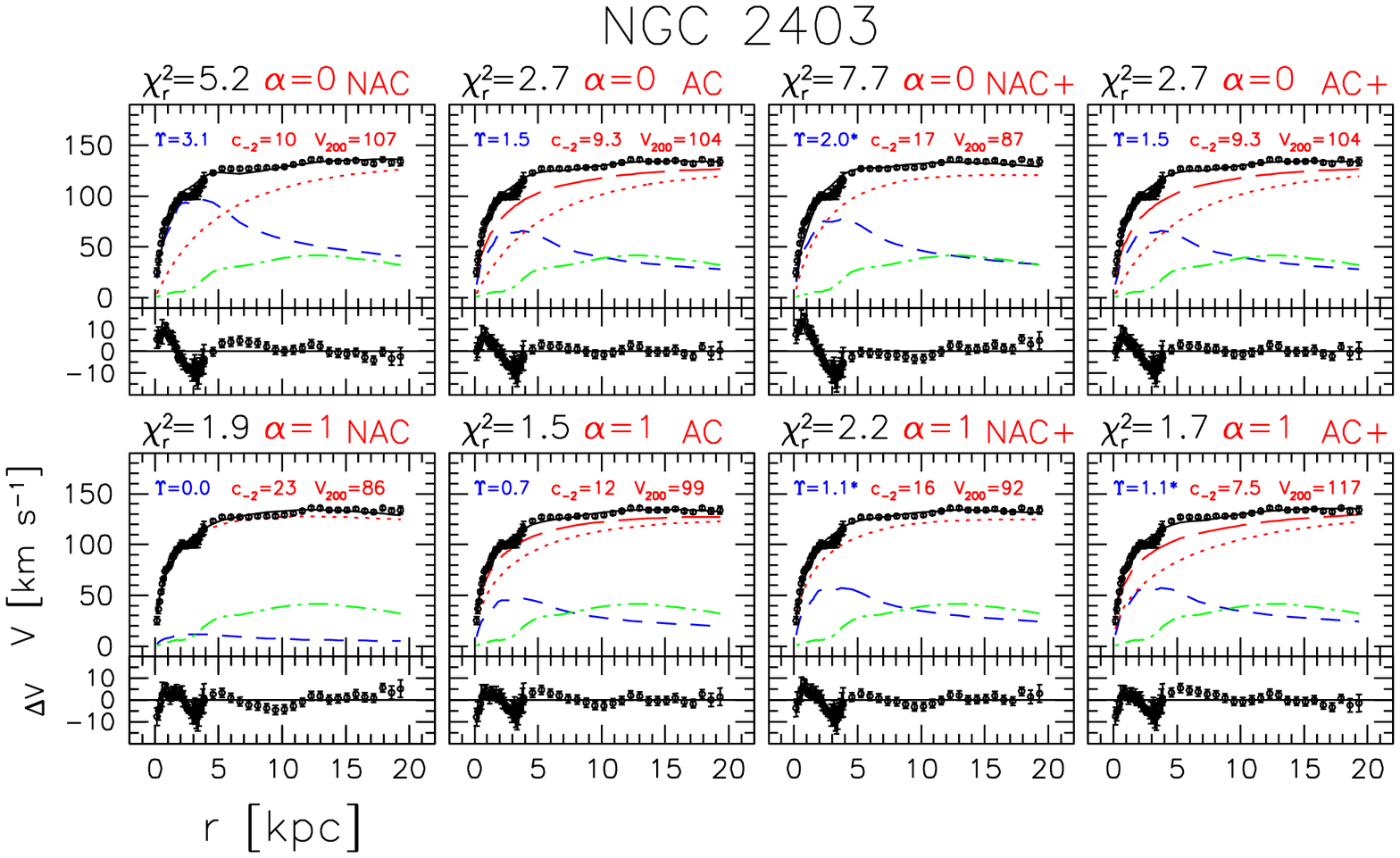}
\includegraphics[bb=  25 360 580 715, width=7.0in]{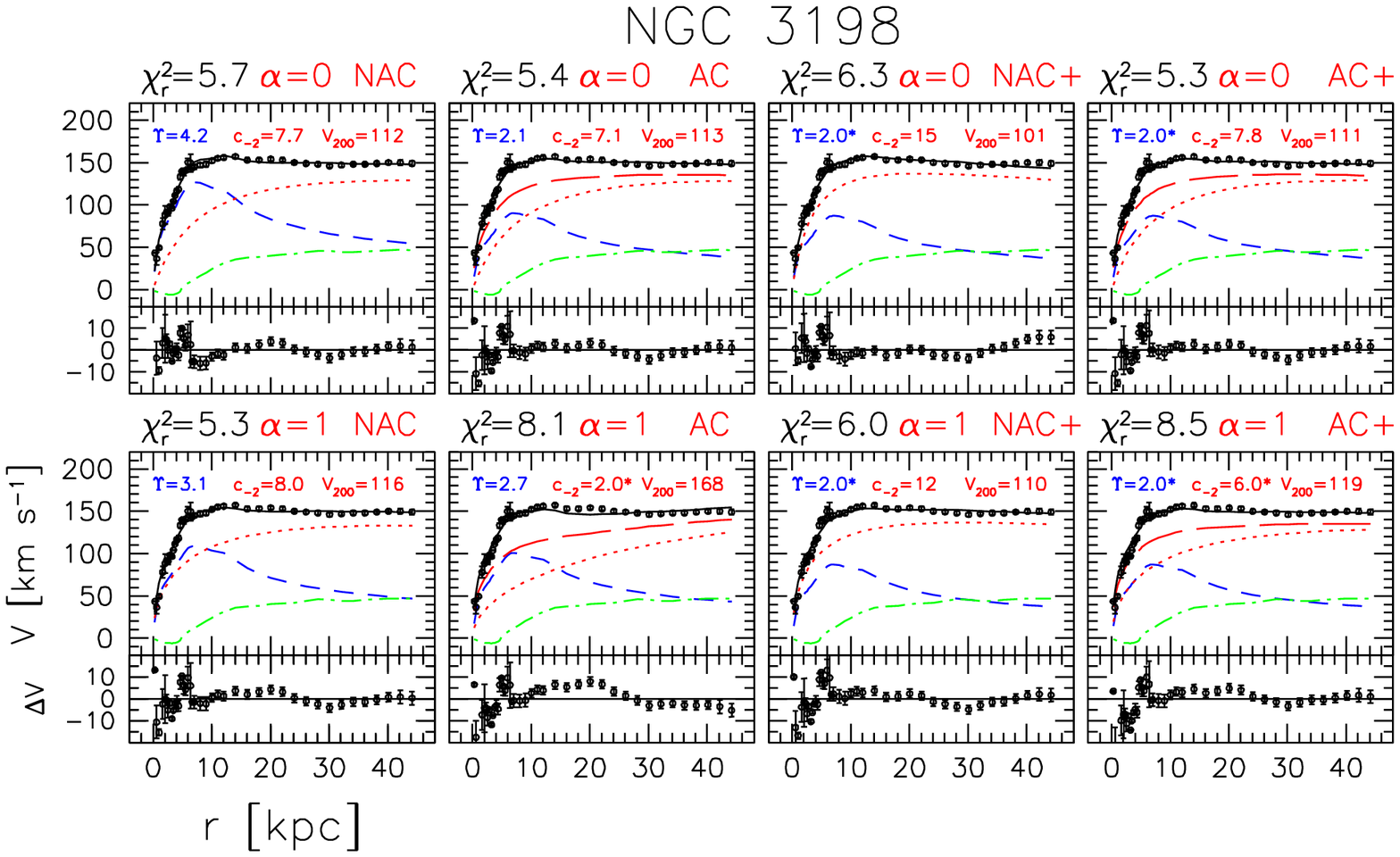}
\figurenum{8} 
\caption{Continued.}
\vspace{0.13in}
\end{center}
\end{figure*} 


\begin{figure*}[t] 
\begin{center}
\figurenum{9} 
\includegraphics[bb=  197 228 510 650, width=6.5in]{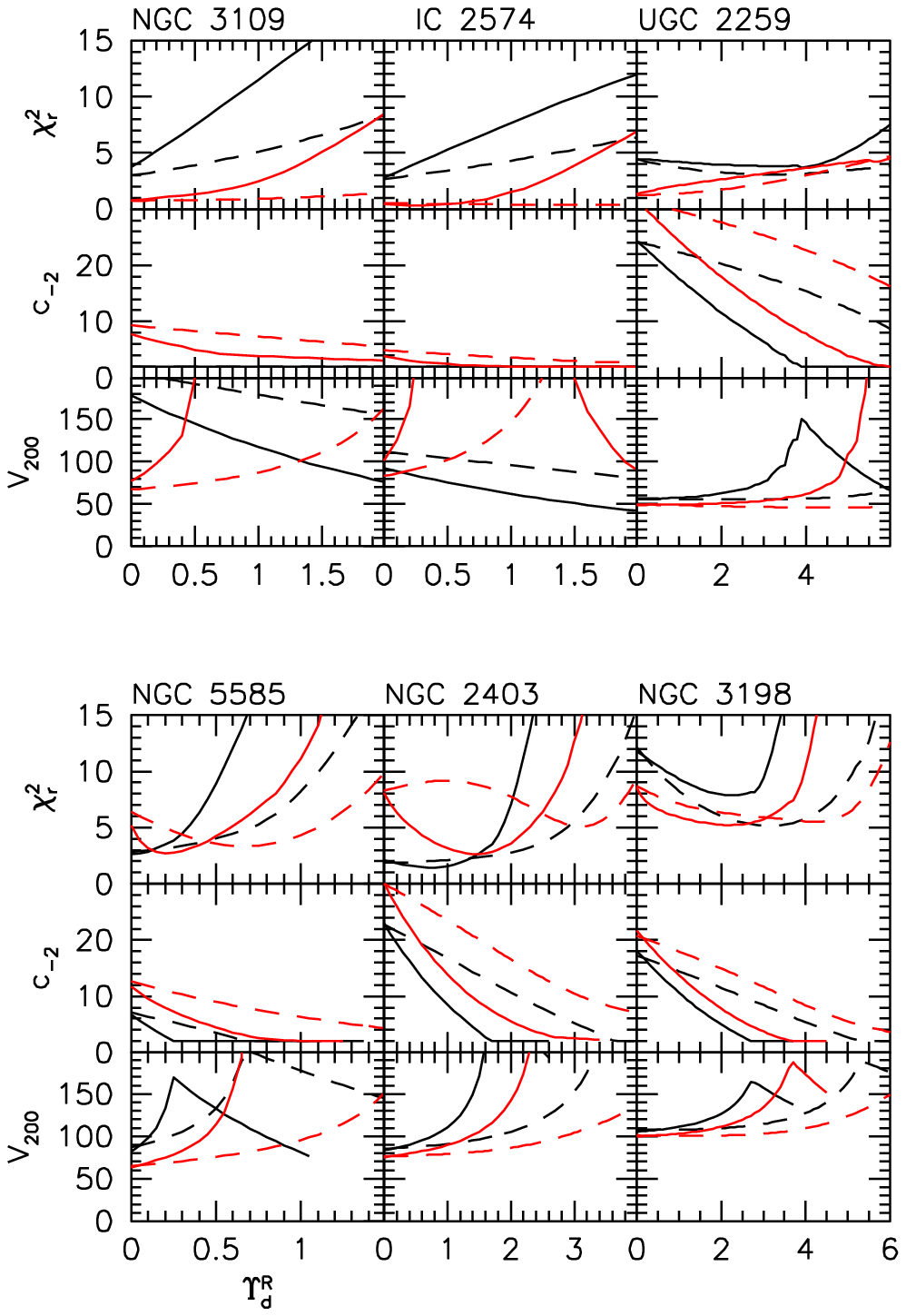}
\caption{Effect on best-fit parameters of adiabatic contraction  
for halos with $\alpha=0$ ({\it red}) and $\alpha=1$ ({\it black}).  
Dashed lines show fits without AC, while solid lines show fits with AC.}
\vspace{-0.2in}
\end{center}
\end{figure*} 

\clearpage

\begin{center} 
\begin{deluxetable*}{ccccccccccc}  
\tablewidth{0pt}
\tabletypesize{\small} 
\tablecolumns{8} 
\tablenum{2} 
\tablecaption{Mass-to-light ratios from constraints} 
\tablehead{
\colhead{Galaxy} & 
\colhead{$\;\;\;\;\;\Upsilon_{\rm d, max}^{R}\;\;\;\;$} & 
\colhead{$\;\;\;\;\;(B-R)_{\rm max}\;\;\;\;$} & 
\colhead{$\;\;\;\;\;(B-R)_{\rm obs}\;\;\;\;\;\;$} & 
\colhead{$\;\;\;\;\;\;\Upsilon_{\rm d, sps}^{R}\;\;\;\;\;\;$} & 
\colhead{$\;\;\;\;\;\Upsilon_{\rm d, sub}^{R}\;\;\;\;$} & 
\colhead{$\;\;\;\;\;\left(V_{\rm disk}/V_{\rm tot}\right)_{2.2}^+\;\;\;\;$} & 
\colhead{$\;\;\;\;\left(V_{\rm disk}/V_{\rm tot}\right)_{\rm max}^+\;\;\;\;$} \\ 
(1) & (2) & (3) & (4) & (5) & (6) & (7) & (8) } 
\startdata 
NGC 3109.........................  
& 1.5  & 1.17 
& 0.9$\pm0.1^a$   &$0.88^{+0.19}_{-0.15}$&    -                
& 0.33 & 0.23 
\\ 
{\phantom{M}} IC 2574.........................  
& 1.5  &  1.17     
& $0.90^b\phantom{\pm.1}$ &  $0.88^{\phantom{\pm0.00}}$&        -             
& 0.41 & 0.24 
 \\  
UGC 2259.........................  
& 8.0  & 2.02 
& 0.9$\pm0.1^a$   &$0.88^{+0.19}_{-0.15}$&    -                
& 0.38 & 0.33 
\\ 
NGC 5585.........................  
& 0.7 & 0.78  
& $0.86^c\phantom{\pm.1}$ & $0.82^{\phantom{+0.00}}_{\phantom{-0.00}}$&  - 
& 0.47 & 0.35 
\\ 
NGC 2403.........................  
& 3.1  & 1.54  
& 1.17$\pm0.15^d$ & $1.5^{+0.5}_{-0.4}$ & $1.1^{+1.1}_{-0.7}$  
& 0.50 & 0.42 
\\  
NGC 3198.........................  
& 4.2  & 1.70  
& 1.17$\pm0.15^d$ & $1.5^{+0.5}_{-0.4}$ & $1.1^{+0.8}_{-0.6}$  
& 0.59 & 0.56
\vspace{1.0mm} 
\enddata 
\tabletypesize{\footnotesize} 
\tablecomments{\footnotesize
Col. (1): Galaxy name.  
Col. (2): Maximum disk $\Upsilon_{\rm d}^R$ from fits including a dark
          halo; for disk only models the $\YdR$ are up to $\sim 25\%$ higher. 
Col. (3): Predicted $B-R$ color for the $\Upsilon_{\rm d}^R$ in
          col. (2) based on SPS models from Bell \& de Jong (2001). 
Col. (4): Observed or expected $B-R$ color. 
Col. (5) predicted $\Upsilon_{\rm d}^R$ for the $B-R$ color given in 
col. (4) using Bell \& de Jong (2001); 
Col. (6) $\Upsilon_{\rm d}^R$ based on the sub-maximal disk constraints from Bottema (1993); 
Col. (7) ratio of disk velocity to total observed rotation velocity at 2.2 disk scale lengths for the best fit $\Upsilon_{\rm d}^R$ with constraints. 
Col. (8) ratio of maximum disk to maximum observed rotation
velocities for the $\Upsilon_{\rm d}^R$ from  Col. (7).} 
\tablenotetext{a}{Expected $B-R$ color for dwarf disk galaxies (e.g. 
 van den Bosch \& Swaters [2001] find 
  $\left\langle B-R \right\rangle =0.87\pm0.09$ 
 for six late-type dwarf galaxies).} 
\tablenotetext{b}{Martimbeau \etal (1994) corrected for Galactic reddening.}
\tablenotetext{c}{C\^{o}t\'{e} \etal (1991) corrected for Galactic reddening.} 
\tablenotetext{d}{Mean $B-R$ color of 40 late-type HSB disk galaxies from MacArthur \etal (2003).} 
\end{deluxetable*} 
\end{center} 

\subsection{Photometry uncertainties} 
In principle, observed surface brightness profiles should be corrected
for projection and  extinction effects to bring the  profiles to their
intrinsic face-on  values. These have opposite effects  on the surface
brightness  profile  (the former  leads to  an  over-estimate of  the
surface brightness, while the latter causes it to be under-estimated).
In   practice,  these   corrections  are fraught   with  significant
uncertainties, especially  for extinction correction.   For simplicity
we only correct for inclination assuming that the disk is
optically  thin and  of zero  thickness.  Thus,  the  observed surface
density will be a factor of $1/\cos(i)$ greater than the face-on case,
where $i$ is the inclination angle. The adopted inclination angles are
given  in  Table~1.  We  correct  for  Galactic  extinction using  the
reddening  values of Schlegel  \etal (1998).   The effect  of internal
dust extinction is to reduce  the observed surface brightness from its
intrinsic  (stellar)  value  and  is  expected  to  have  some  radial
dependence. Thus, the simple dust-free  model that we adopt, all other
variables being equal, gives an {\it upper limit} to the $\YdR$.


\section{Rotation Curve Fits} 
As a  result of certain  covariances between model parameters,  we fit
for \ccm2  and \v200  on a grid  of \YdR  and $\alpha$ to  prevent our
fitting routine from getting trapped  in local minima.  The results of
these fits are shown in Figure~7, where we plot the reduced \chisq and
best-fit \ccm2  and \v200  against $\alpha$ for  a range of  \YdR from
zero to maximum disks.   Examples of rotation curve decompositions for
$\alpha=0$ and $\alpha=1$ halos are shown in Fig.~8.  Clearly, without
constraints  the models  are highly  degenerate, with  acceptable fits
being possible covering a  wide, ill-constrained range of $\alpha$ and
$\YdR$.

\subsection{Effect of adiabatic contraction} 
In Figure~9 we  plot the \chisqr and best fit  \ccm2 and \v200 versus\
\YdR,  for  $\alpha=0$ and  $\alpha=1$  halos.  Adiabatic  contraction
causes the rotation velocity of  the halo to rise, especially at small
radii.  Thus,  to obtain a  comparable circular velocity  profile, the
concentration  of the  pre-contracted  halo must  be  lower.  As  \YdR
increases, this effect becomes  more significant and can substantially
alter the halo  parameters, even for sub-maximal disks.   For close to
maximum disks acceptable fits  cannot be obtained without breaking the
weak  constraint: $\ccm2  \ge 2$.   Even after  exclusion of  the high
\vdisk/\vtot solutions, the disk-halo and cusp-core degeneracies still
remain.   In  general, the  best-fit  \chisqr  values  are lower  with
adiabatic contraction than without.  In  cases in which a maximum disk
is the preferred fit  without adiabatic contraction, the best-fit \YdR
and \chisqr are both lower with adiabatic contraction.

\begin{figure*}
\vspace{0.1in}
\begin{center}
\figurenum{10} 
\includegraphics[bb=  125 380 630 685, width=7.6in]{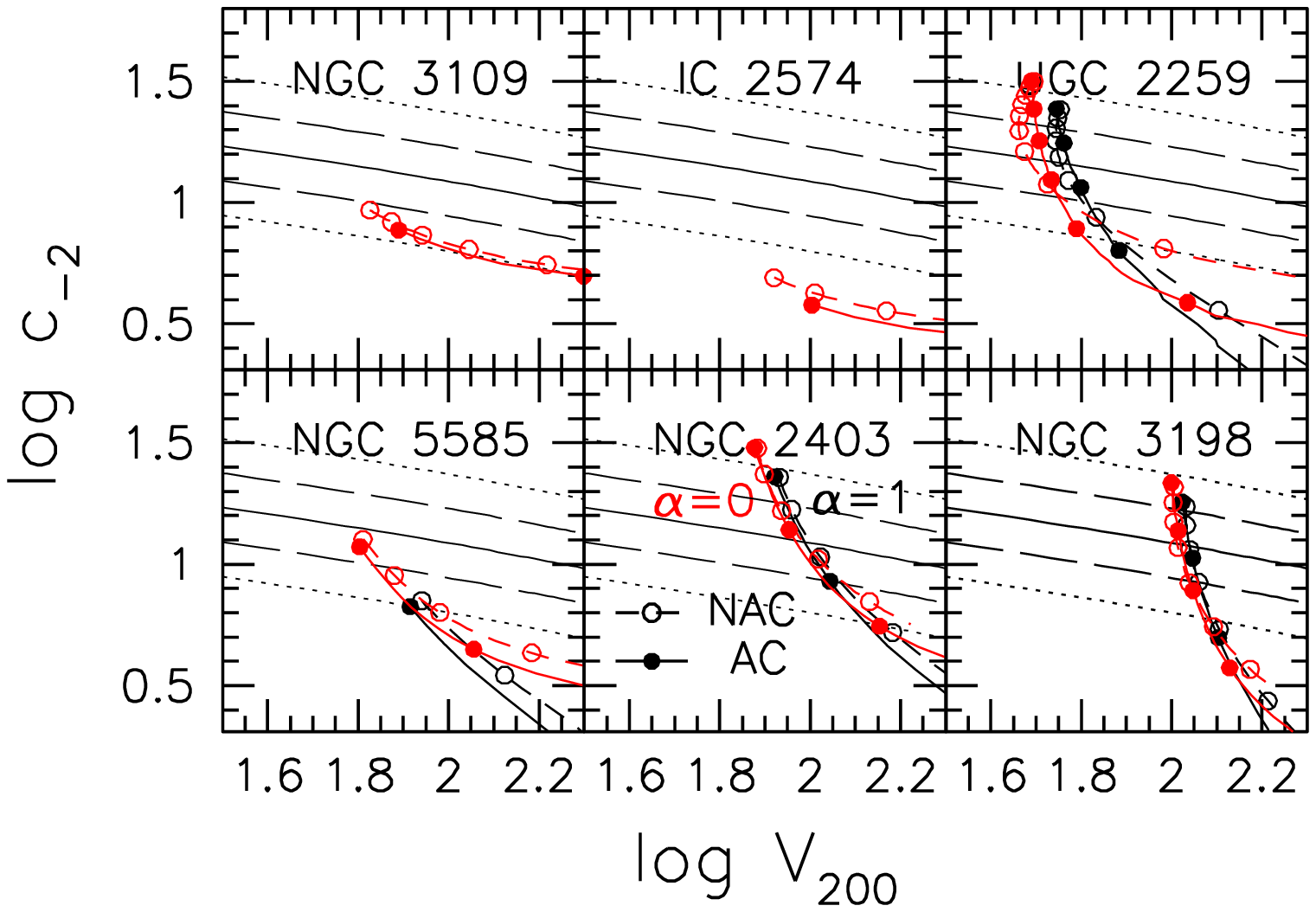}
\caption{Comparison between our fitted $c_{-2}$ and $V_{200}$ against 
the model from Bullock \etal (2001). The mean is given by the solid line, 
while the 1 and 2 $\sigma$ deviations are given by the long-dashed and
dotted lines,  
respectively. Fits with $\alpha=0$ are in black, while fits with $\alpha=1$  
are in red. Fits with adiabatic contraction (AC) are given by filled circles 
and solid lines, while fits without AC (NAC) are given by open circles and  
dotted lines.  For NGC 3109, IC 2574, and NGC 5585 the symbols are at 
$\Upsilon_{\rm d}^R$ intervals of 0.5, while for UGC 2259, NGC 2403, and 
NGC 3198 the interval is 1.0.} 
\end{center}
\end{figure*} 


\subsection{Fits with constraints} 
{\it  SPS models.}---Table  2  gives the  observed  or expected  $B-R$
colors    and     the    predicted    \YdR     using    the    formula
$\log(\YdR)=-0.820+0.851(B-R)$ (Table  1, Bell  \& de Jong  2001).  We
find  that the  galaxy colors  and the  SPS-predicted \YdR  values are
significantly  lower  than  the  rotation curve-derived  maximal  disk
values for  five out  of six galaxies,  the exception being  NGC 5585,
although  the latter  is dark  matter dominated  at larger  radii.  In
terms of velocities  $\left(V_{\rm disk}/V_{\rm tot}\right)_{2.2} \lta
0.6$  for  all  six  galaxies.   As  we  have  not  included  internal
extinction  effects,  the  intrinsic  colors  will be  bluer  and  the
resulting \YdR and disk velocity fractions will be lower.  The maximal
disk \YdR for NGC 5585  is inconsistent with $\alpha\simeq1$ halos and
cannot be reconciled with any halo that is adiabatically contracted.
 
{\it   Sub-maximal  disks.}---NGC   3198  has   a   measured  velocity
dispersion,  ${\left\langle V_z^2 \right\rangle}^{1/2}_{R=0}=40\pm0.7$
(Bottema  1993), implying $\left(V_{\rm  disk}/V_{\rm obs}\right)_{\rm
max}  =0.43\pm0.15$.   For NGC  2403  we  apply  the sub-maximal  disk
constraint (Eq.~10).  The associated \YdR are given in Table 2 and are
consistent  with  the  expected  \YdR  from SPS  models.   {\it  These
constraints  rule out  zero and  maximal disks  but do  not  break the
disk-halo degeneracy  for NGC 2403 and  NGC 3198, even  when the halos
are adiabatically contracted}.

{\it The $\ccm2  - \v200 $ comparison.}---In Figure~10  we compare the
fitted \ccm2  and \v200 for  $\alpha=1$ and $\alpha=0$ halo  fits with
the \ccm2 $-$ \v200 model of Bullock \etal (2001).  This shows that as
\YdR increases, the fitted \ccm2 decreases and \v200 increases.  For a
given {\YdR}, fits  with $\alpha=0$ have larger \ccm2  and lower \v200
than  fits with $\alpha=1$,  but the  differences are  not significant
enough to  distinguish between  $\alpha=0$ and $\alpha=1$.   Fits with
adiabatic  contraction follow  the same  path in  the $c_{-2}-V_{200}$
space as  those without, such that for  a given \YdR the  \ccm2 of the
adiabatic contraction  halos are lower.   (Note that the  $c_{-2}$ and
$V_{200}$ values of the contracted halos are those of the {\it initial
} pre-contracted halo.)  For NGC 2403, NGC 3198, and UGC 2259 the \YdR
values from  SPS models  fall within $1\sigma$  of the mean  \ccm2 for
$\alpha=1$ halos, while  for IC 2574, NGC 3109, and  NGC 5585 the \YdR
from SPS models require very low $\ccm2 $.
 
{\it Lensing-TF.}---This constraint  (Eq.~14) corresponds to $V_{200}$
=  $76^{+21}$  and  $87^{+25}$  \kms   for  NGC  2403  and  NGC  3198,
respectively.  Note that  there is no lower limit  on $V_{200}$.  From
Fig.~7, we see  that for NGC 2403 the lowest \v200  = 75 $\kms$, while
for NGC  3198 the lowest \v200  = 100 $\kms$. These  values occur when
\ccm2 is highest,  which corresponds to $\alpha=0$ and  $\YdR =0$.  As
$\alpha$ and  \YdR increase, \ccm2  decreases and \v200  increases, so
that the highest  (and thus excluded) values of  \v200 occur for cuspy
halos with close  to maximal disks.  Thus, this  constraint favors low
$\alpha$ and \YdR fits.

{\it All  constraints.}---Our full set of constraints  consist of \YdR
from  SPS models,  $6\le c_{-2}\le  30$, and  $\v200 \le  \vmax$.  The
rotation curves for NGC 3109 and  IC 2574 are still rising at the last
data  point, so  we adopt  \vmax =  100, consistent  with other 
dwarf 
galaxies.  In Table~3  we give the ranges in  $\alpha$ that are within
3.5 of  the best-fit  $\chisqr$. While we  cannot place  a statistical
confidence  on  these  ranges,  they  seem to  encompass  all  of  our
acceptable  fits.  Clearly,  even  with the  full  set of  constraints
significant degeneracies remain, often  with both $\alpha\simeq 0$ and
$\alpha\simeq  1$ halos  providing reasonable  fits.   The differences
between   fits  with   and  without   adiabatic  contraction   can  be
substantial.  Given  that other  evolutionary processes exist  and are
not  clearly  understood,  there  remains significant  uncertainty  in
determining $\alpha$.

\begin{center} 
\begin{deluxetable}{ccccc}
\tablewidth{0pt} 
\tabletypesize{\small}
\tablecolumns{5} 
\tablenum{3} 
\tablecaption{Range    and    best-fit   values    of    $\alpha$.}
\tablehead{
Galaxy & 
$\;\;\;\;\alpha_{NAC}\;\;\;$ & 
$\;\;\;\;\alpha_{AC}\;\;\;$ & 
$\;\;\;\;\alpha_{NAC+}\;\;\;$ &
$\;\;\;\;\alpha_{AC+}$ \\ 
(1) & (2) & (3) & (4) & (5)} \startdata
NGC 3109.................  
& $0.0^{\ 1.1}_{\ 0.0}$  
& $0.0^{\ 1.1}_{\ 0.0}$  
& $0.0^{\ 0.8}_{\ 0.0}$ 
& $0.0^{\ 0.3}_{\ 0.0}$ \\ 
$\phantom{II}$IC 2574.................  
& $0.0^{\ 1.1}_{\ 0.0}$  
& $0.0^{\ 1.1}_{\ 0.0}$  
& $0.0^{\ 0.4}_{\ 0.0}$ 
& $-\phantom{0.0}$ \\ 
UGC 2259.................  
& $0.0^{\ 1.5}_{\ 0.0}$  
& $0.0^{\ 1.3}_{\ 0.0}$  
& $0.0^{\ 1.1}_{\ 0.0}$  
& $0.0^{\ 1.1}_{\ 0.0}$ \\ 
NGC 5585.................  
& $1.0^{\ 1.4}_{\ 0.0}$  
& $1.0^{\ 1.4}_{\ 0.0}$  
& $0.5^{\ 0.9}_{\ 0.0}$  
& $-{\phantom{0.0}}$\\ 
NGC 2403.................  
& $1.2^{\ 1.5}_{\ 0.5}$  
& $0.9^{\ 1.5}_{\ 0.0}$  
& $1.2^{\ 1.5}_{\ 0.6}$  
& $1.0^{\ 1.3}_{\ 0.0}$\\ 
NGC 3198.................  
& $0.9^{\ 1.4}_{\ 0.0}$  
& $0.0^{\ 1.1}_{\ 0.0}$  
& $0.8^{\ 1.3}_{\ 0.0}$  
& $0.0^{\ 1.0}_{\ 0.0}$
\vspace{0.1cm}
\enddata 
\tabletypesize{\footnotesize}
\tablecomments{\footnotesize
Col. (1): galaxy name.   
Col. (2): best fit and upper and lower limits based on a $\Delta \chi^2_r=3.5$ for fits without adiabatic contraction (AC) or constraints (+).  
Col. (3): Same as col. (2), but for fits with AC. 
Col. (4): Same as col. (4), but for fits with constraints. 
Col. (5): Same as col. (2), but for fits with AC and constraints. 
} 
\end{deluxetable} 
\end{center}


\section{Other uncertainties in mass models} 
We now  discuss the sensitivity  of our results to  several parameters
that have so  far been kept fixed. These  include the minimum rotation
curve  error  values, $dv_{\rm  min}$;  the  distance,  $D$; the  disk
thickness,  $\qd$; and the  halo flattening,  $\qh$.  Results  for the
tests  we  perform below  are  shown  in  Fig.~11 for  $\alpha=0$  and
$\alpha=1$  halos  without  adiabatic  contraction.  The  results  are
similar with adiabatic contraction, but the range in \YdR is lower.

{\it  Minimum rotation  curve  error values.}---We  use minimum  error
values for each velocity measurement at  a given radius of 2, 3, and 4
$\kms$. Larger  error values have  the effect of lowering  the \chisqr
and make it harder to distinguish between models.
 
{\it  Distance.}---We show  the effect  of a  30\% distance  error, in
order  to account  for uncertainties  in distance  estimation  such as
peculiar velocities.  The mass of the stars and gas is proportional to
distance, so for larger distances and a given \vdisk \YdR is lower.
 
{\it Disk thickness.}---We  consider the values of $q_0=0$  for a thin
disk, the  commonly-used $q_0=1/6$, and  $q_0=0.25$.  The effect  of a
thick disk  is to  smooth out the  features of the  surface brightness
profile and lower the amplitude of the rotation curve (Fig.~12).  This
results in a higher maximum disk \YdR but does not significantly alter
the relative goodness of fit between $\alpha=0$ and $\alpha=1$ halos.

{\it  Halo  flattening.}---We  compute  fits  for  oblate  halos  with
$q=0.25$ and $q=0.5$  and a prolate halo with  $q=1.5$.  The effect of
halo flattening on  \chisqr is very small, and  is only noticeable for
low \YdR  fits because the  halo parameters adjust; as  $q$ decreases,
\ccm2 decreases  and \v200 stays roughly constant.   Because we define
$R_{200}$  to  be independent  of  $q$,  then  for a  given  $r_{-2}$,
$V_{200}$  increases as  $q$  decreases.  Therefore,  if $V_{200}$  is
constant,  $c_{-2}$ will  decrease as  $q$ decreases.   Thus, although
halo  flattening does  not  alter the  $\chi^2_r$,  it contributes  an
additional $\sim 20\%$ uncertainty to $c_{-2}$.


\begin{figure*}
\begin{center} 
\includegraphics[bb=  20 320 570 650, width=7.0in]{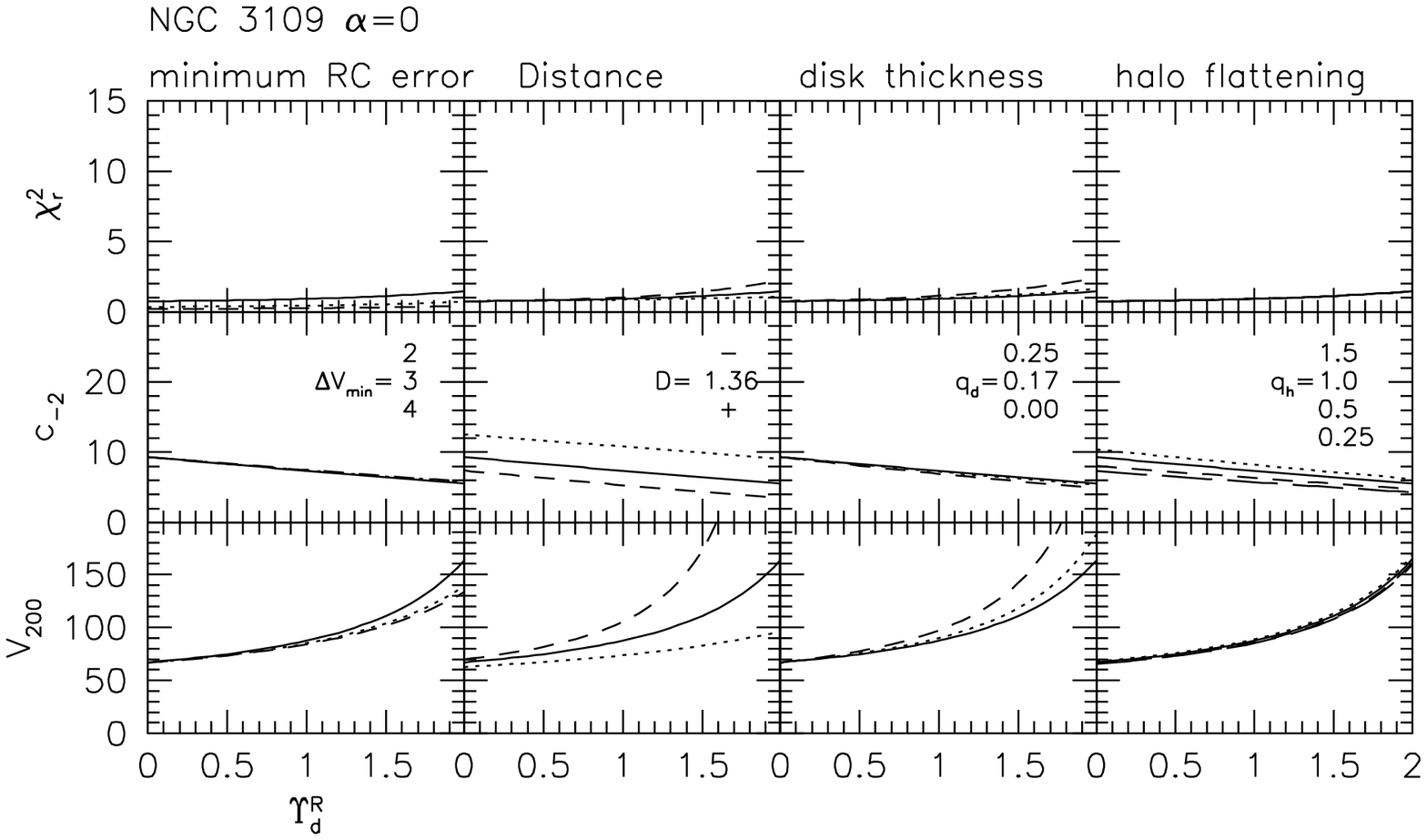}
\includegraphics[bb=  20 320 570 650, width=7.0in]{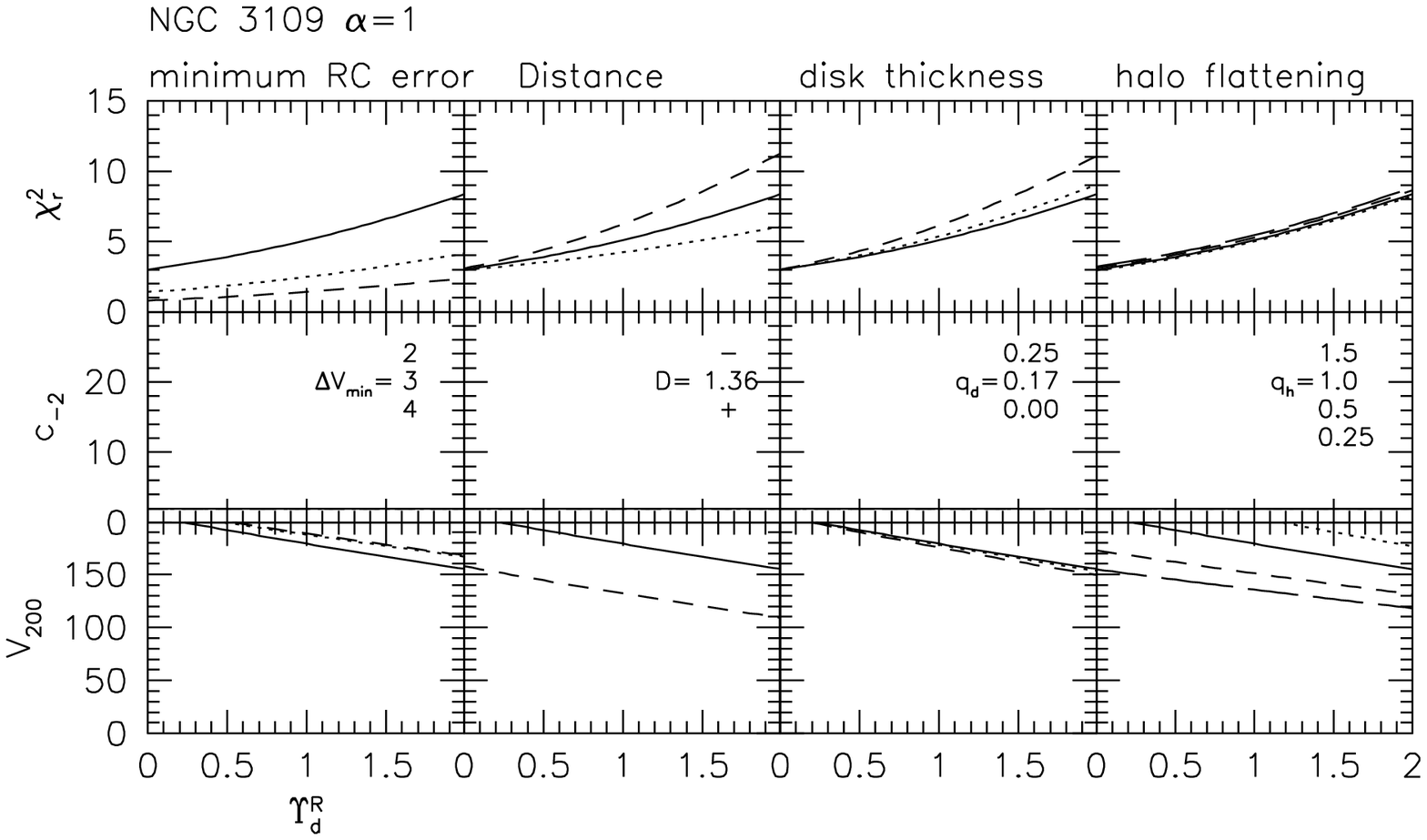}
\figurenum{11} 
\caption{Effect of minimum rotation curve errors, distance, disk thickness, 
and halo flattening for $\alpha=$0 and 1 halos for all six galaxies. 
Minimum RC errors: 2 ({\it solid line}), 3 ({\it dotted line}), 4
  ({\it dashed line} ); Distance: adopted ({\it solid line}), 
-30\% ({\it dotted line }), +30\% ({\it dashed line}); disk thickness: 
0.25 ({\it solid line}), 0.17 ({\it dotted line}), 0.00 ({\it dashed line}); 
halo flattening: 1.00 ({\it solid line}), 1.50 ({\it dotted line}), 
0.50 ({\it dashed line}), 0.25 ({\it long-dashed line}).} 
\end{center}
\end{figure*} 
 
\begin{figure*}
\begin{center} 
\includegraphics[bb=  20 320 570 650, width=7.0in]{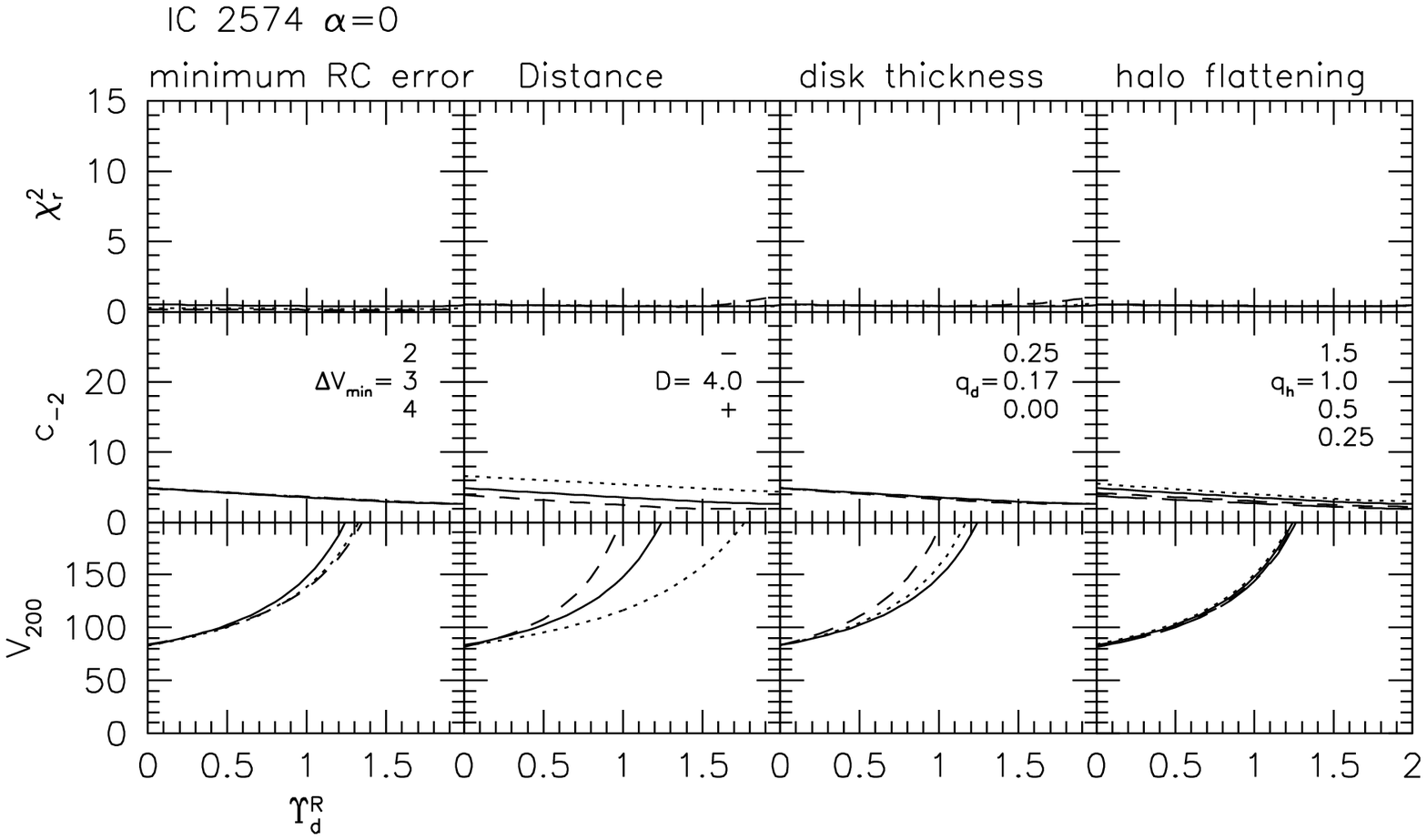}
\includegraphics[bb=  20 320 570 650, width=7.0in]{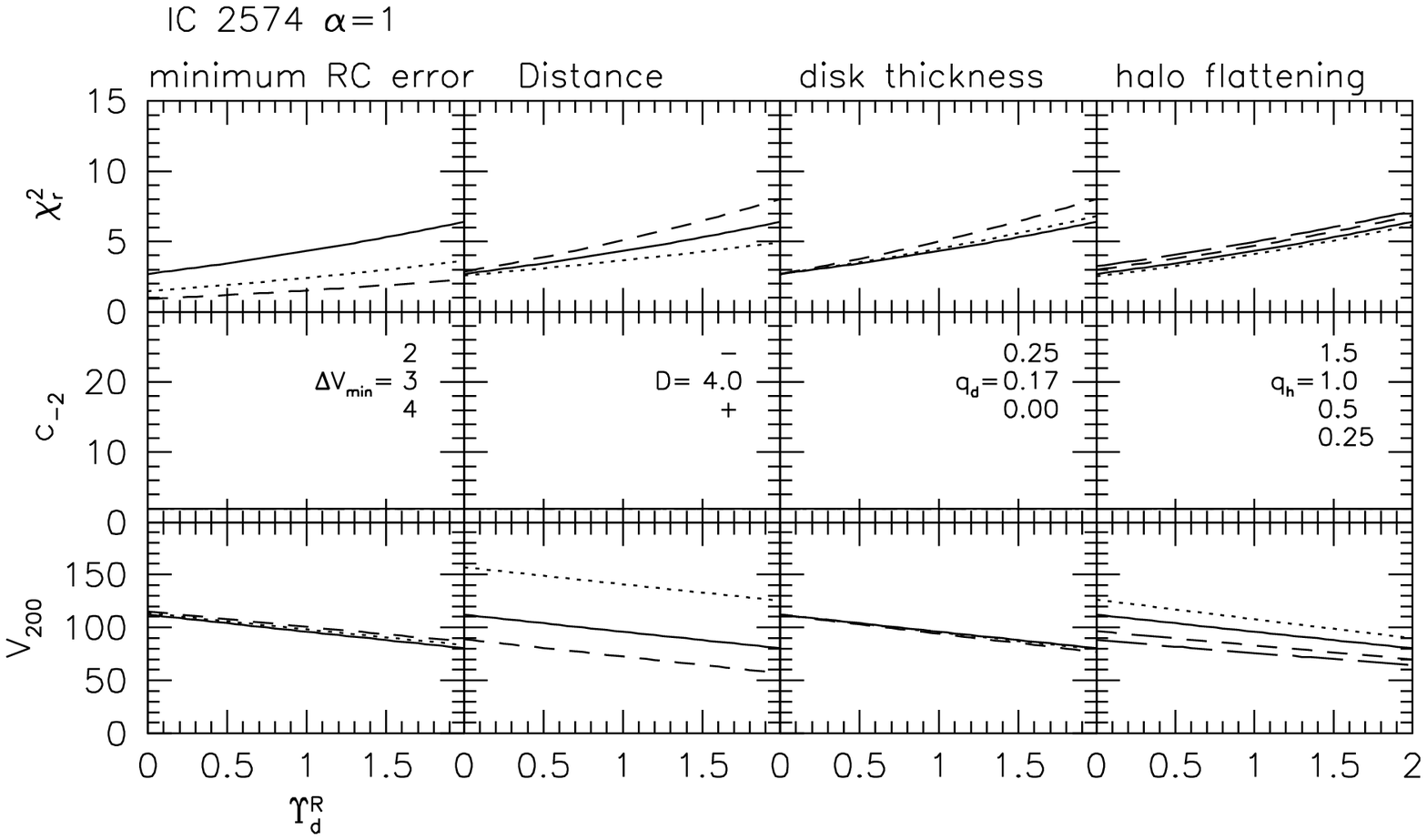}
\figurenum{11}
\caption{Continued.}
\end{center}
\end{figure*} 
 
\begin{figure*}
\begin{center} 
\includegraphics[bb=  20 320 570 650, width=7.0in]{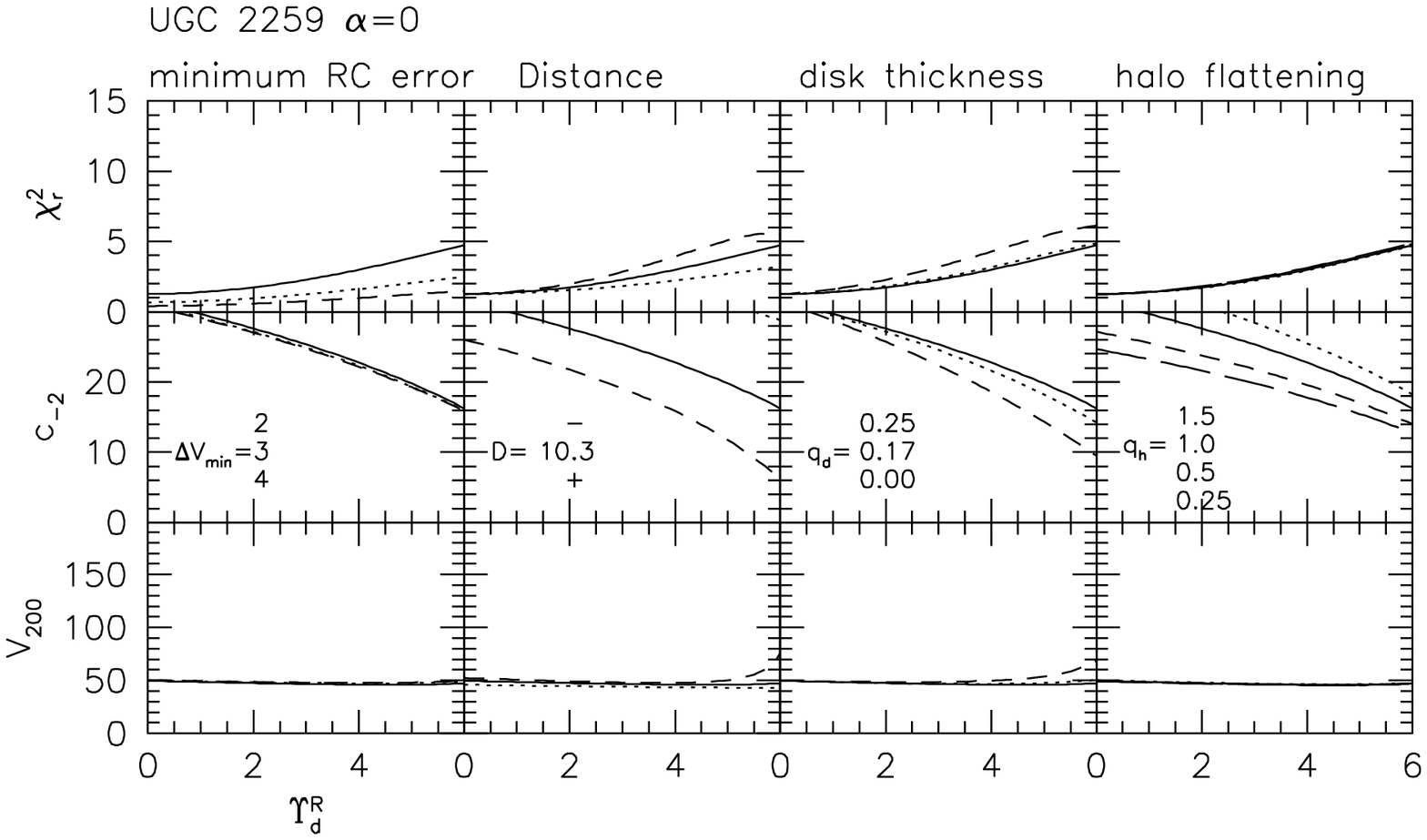}
\includegraphics[bb=  20 320 570 650, width=7.0in]{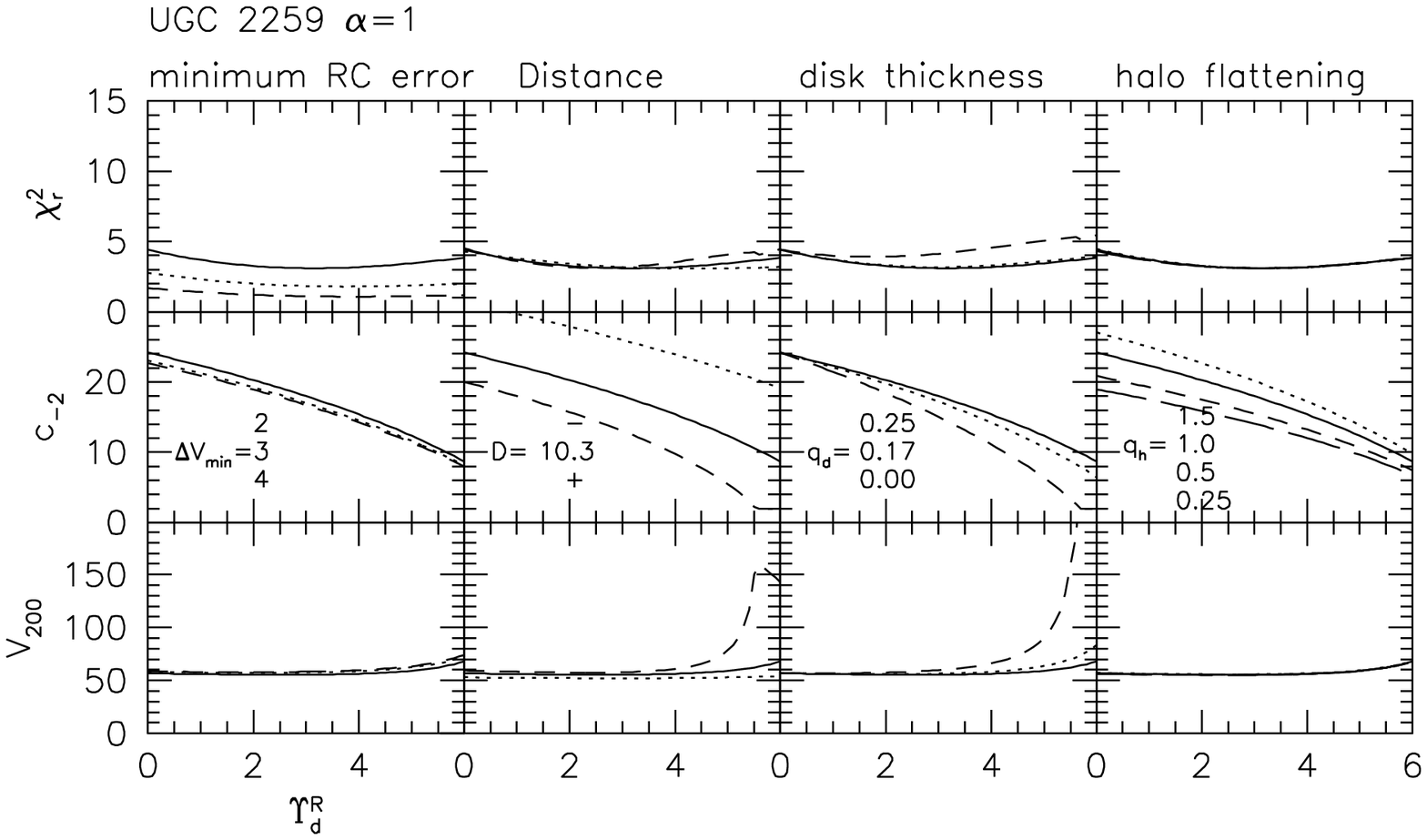}
\figurenum{11}
\caption{Continued.} 
\end{center}
\end{figure*}

\begin{figure*}
\begin{center} 
\includegraphics[bb=  20 320 570 650, width=7.0in]{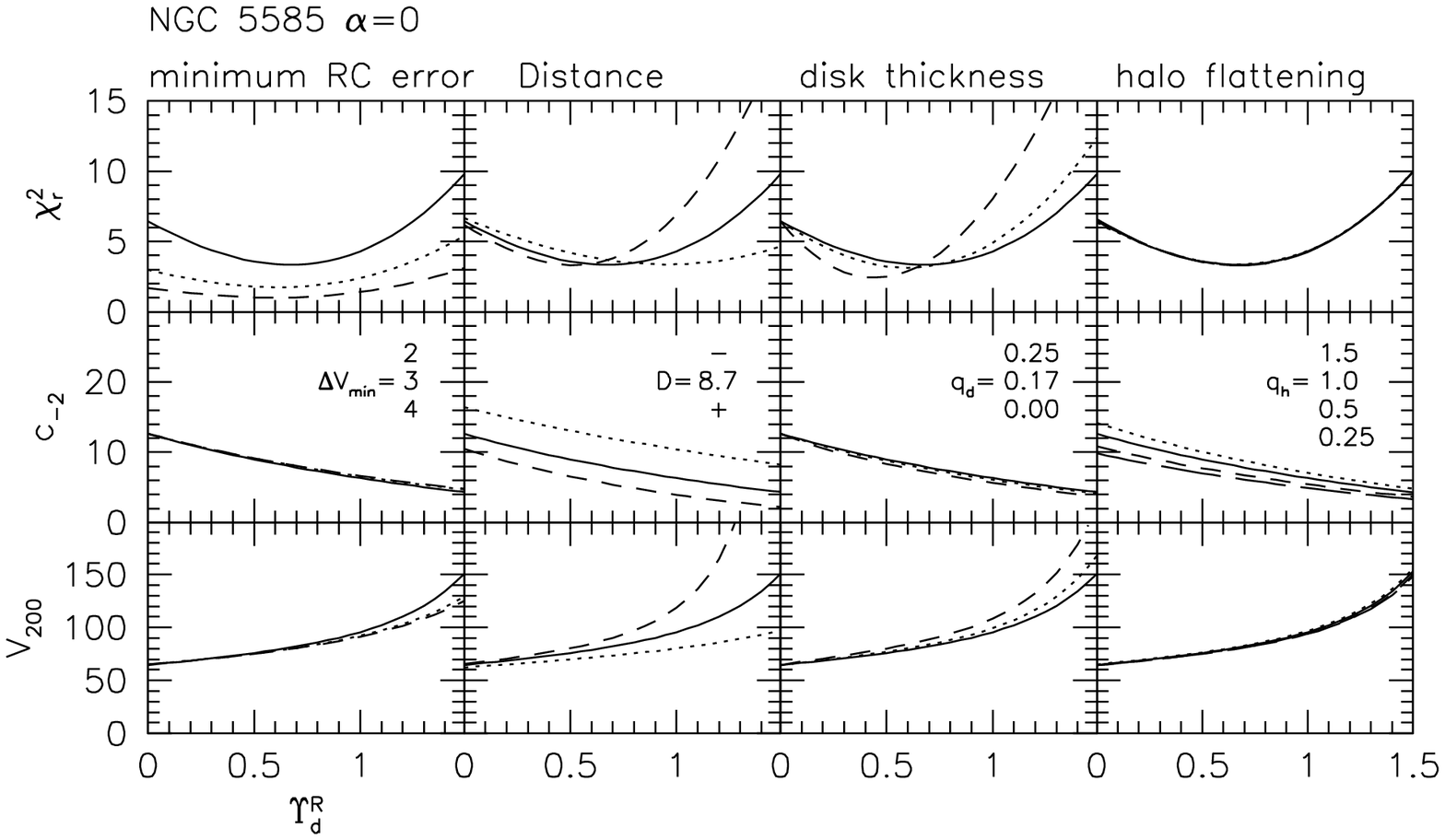}
\includegraphics[bb=  20 320 570 650, width=7.0in]{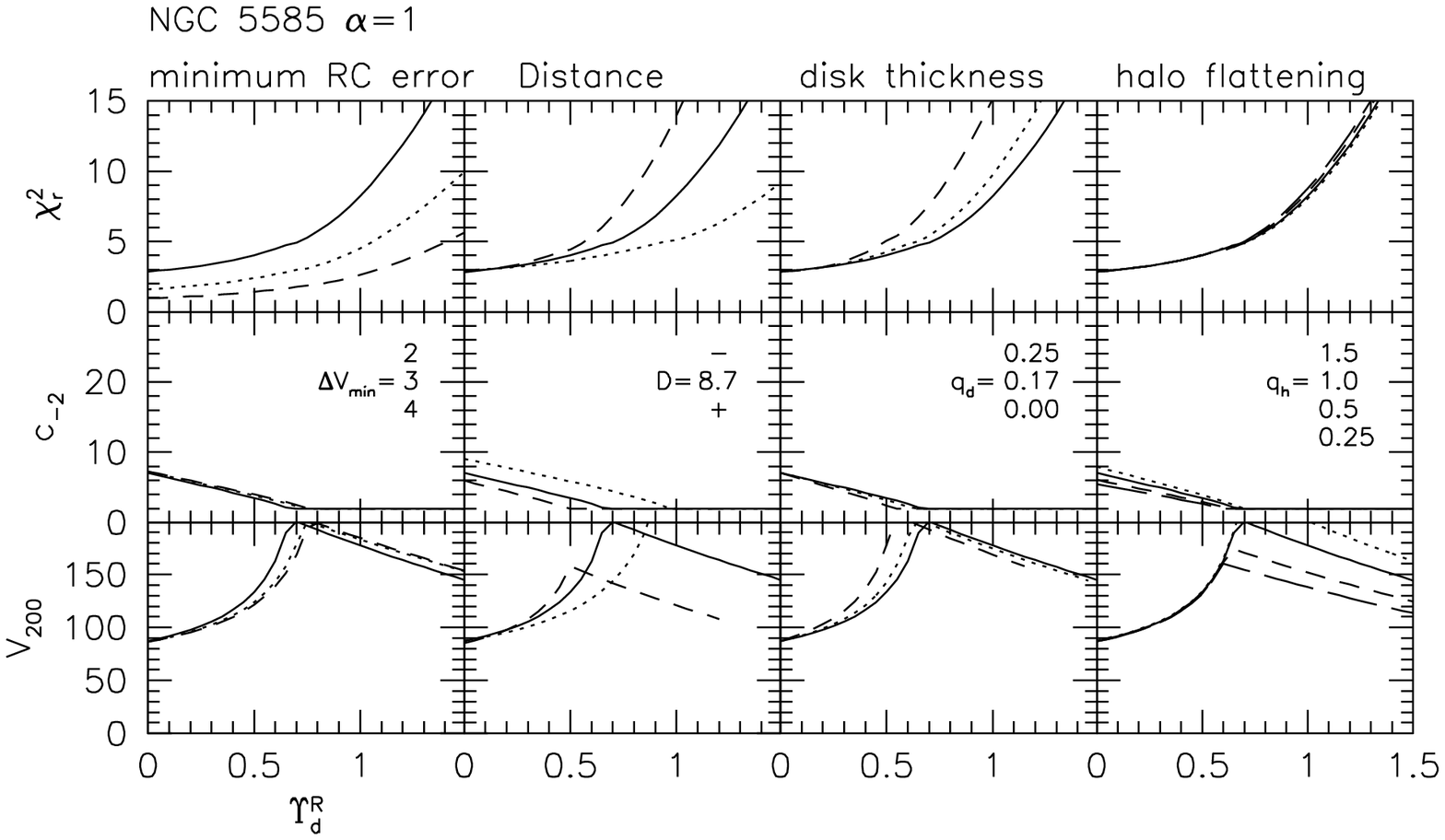}
\figurenum{11}
\caption{Continued.}
\end{center} 
\end{figure*}

\begin{figure*}
\begin{center} 
\includegraphics[bb=  20 320 570 650, width=7.0in]{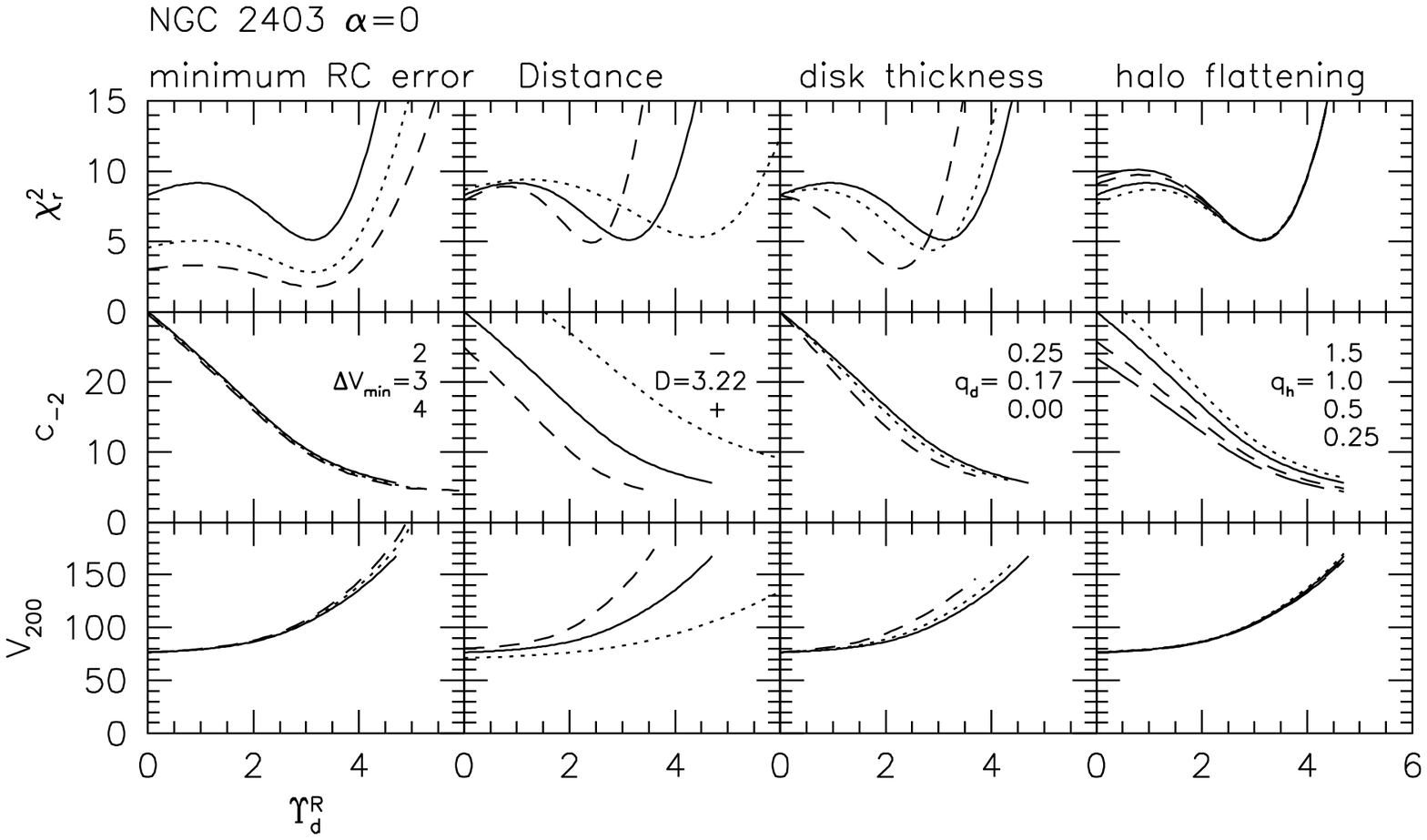}
\includegraphics[bb=  20 320 570 650, width=7.0in]{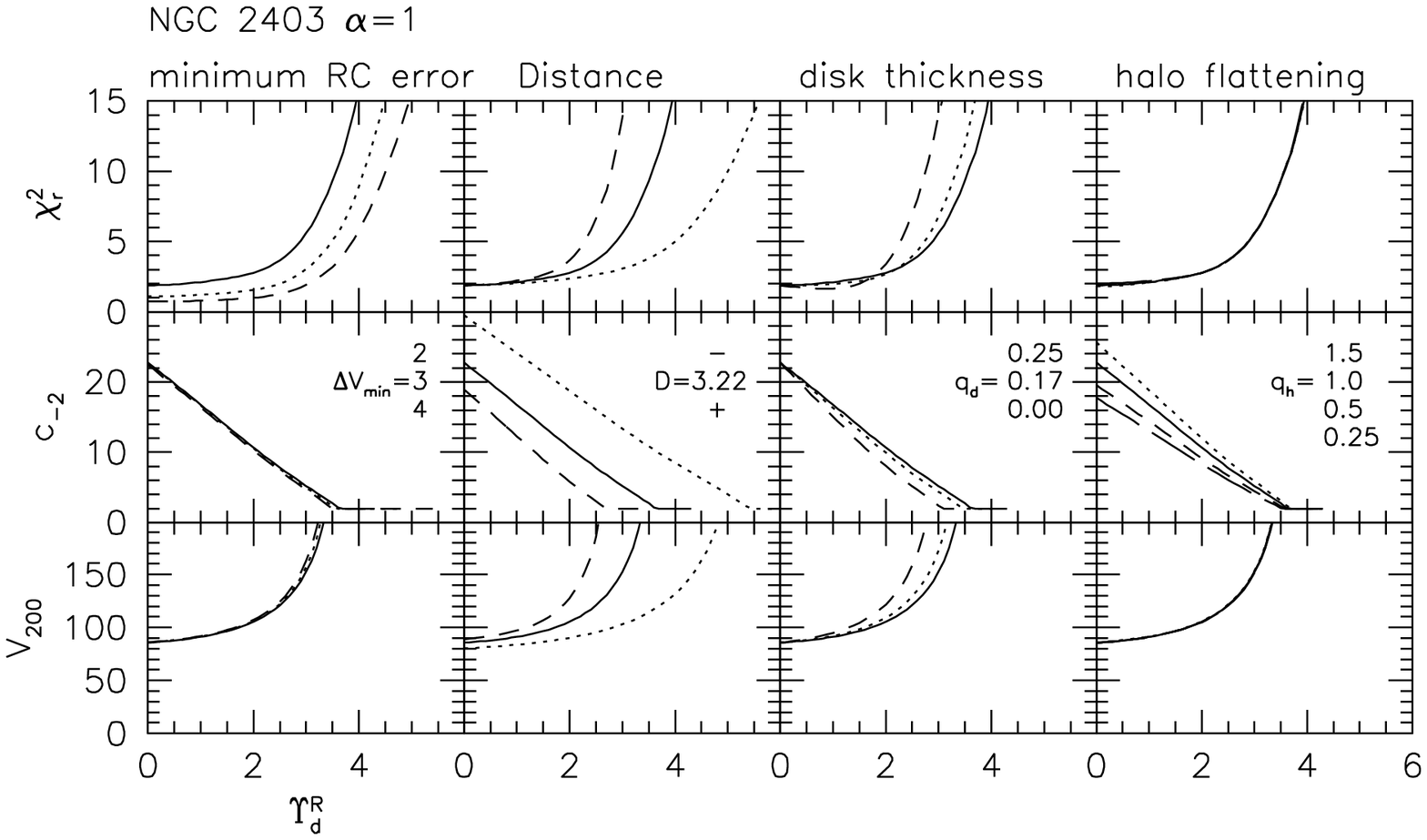}
\figurenum{11}
\caption{Continued.}
\end{center} 
\end{figure*}

\begin{figure*}
\begin{center} 
\includegraphics[bb=  20 320 570 650, width=7.0in]{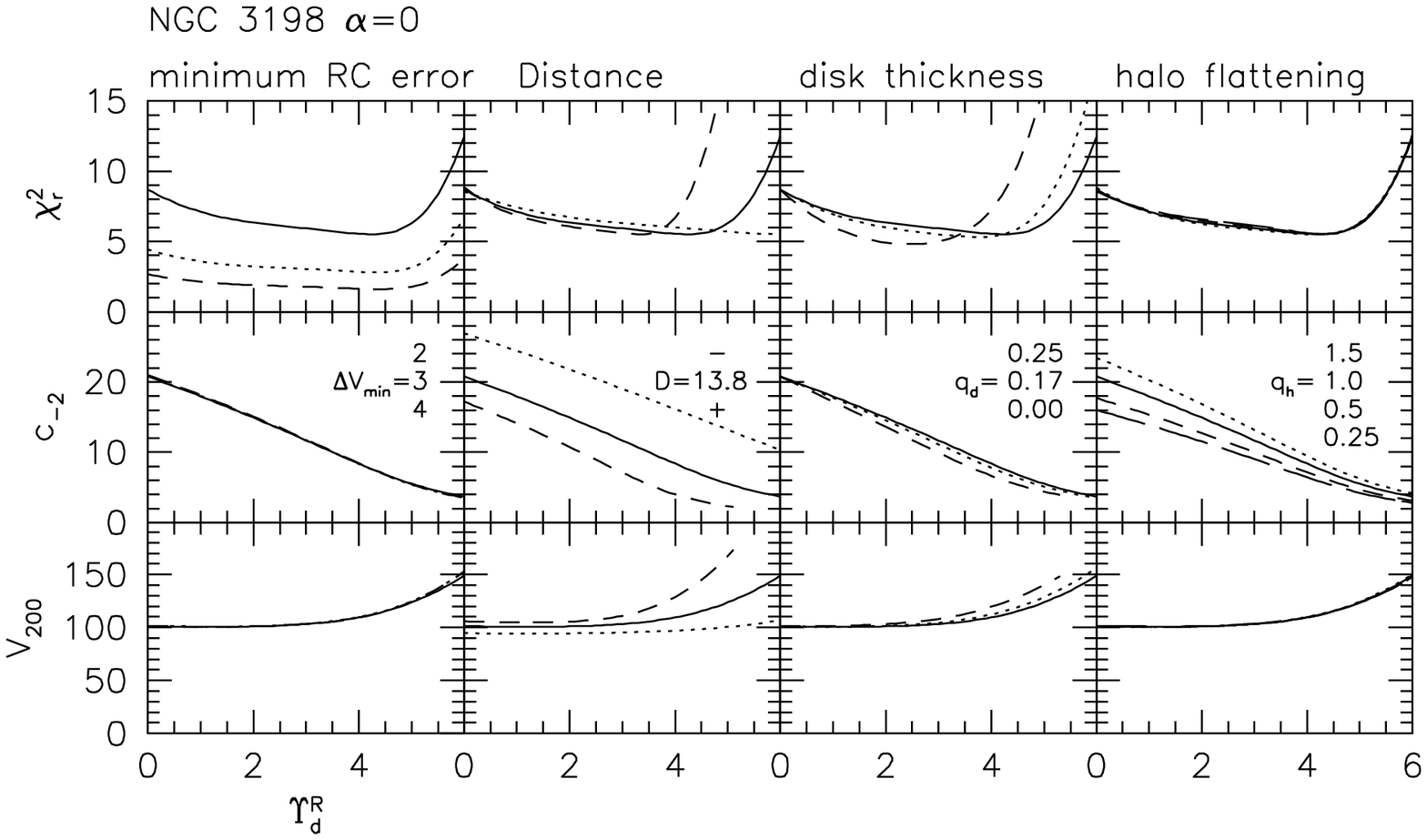}
\includegraphics[bb=  20 320 570 650, width=7.0in]{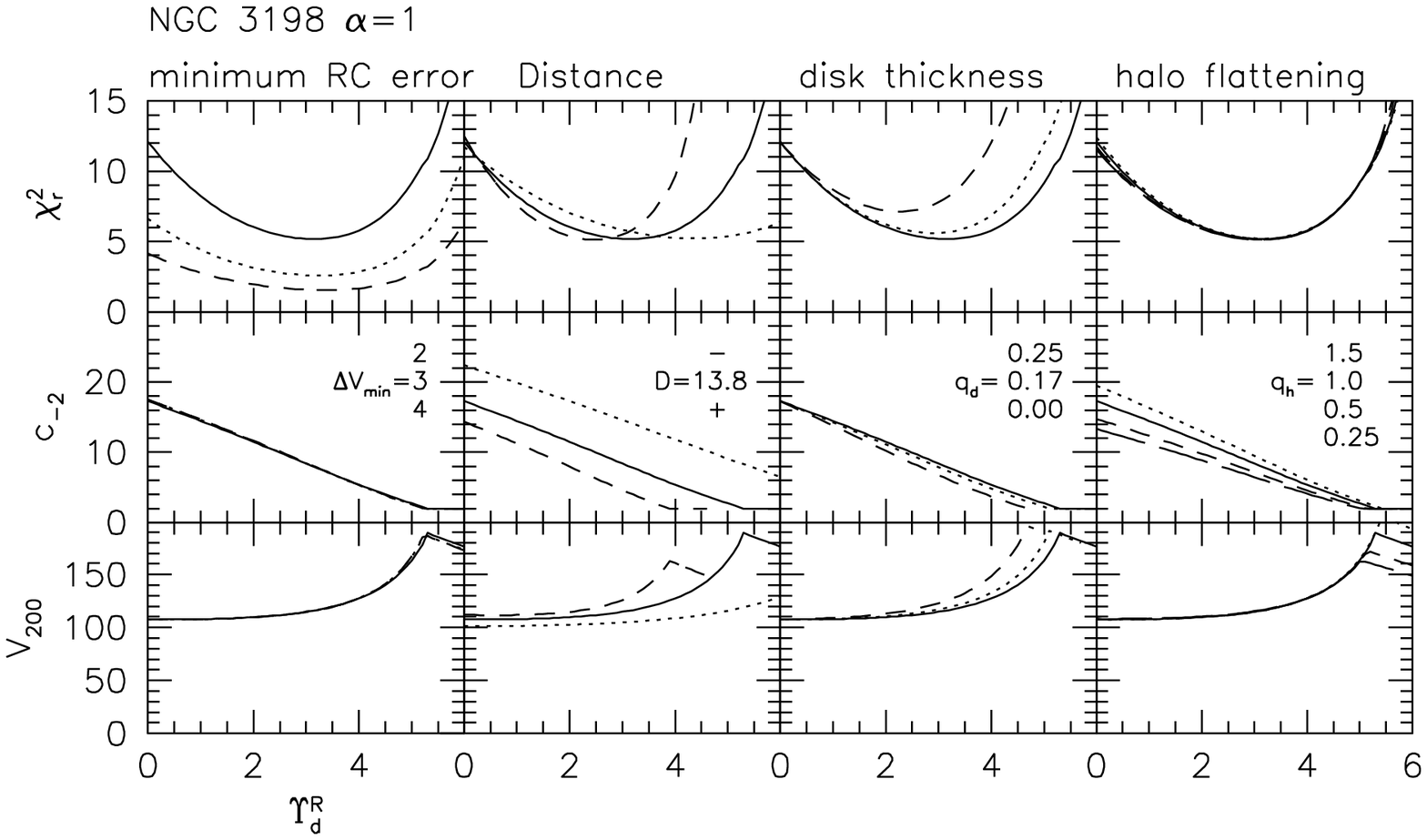}
\figurenum{11}
\caption{Continued.}
\end{center} 
\end{figure*}

\begin{figure*}[t]
\begin{center}  
\figurenum{12} 
\includegraphics[bb=  60 200 590 560, width=7.2in]{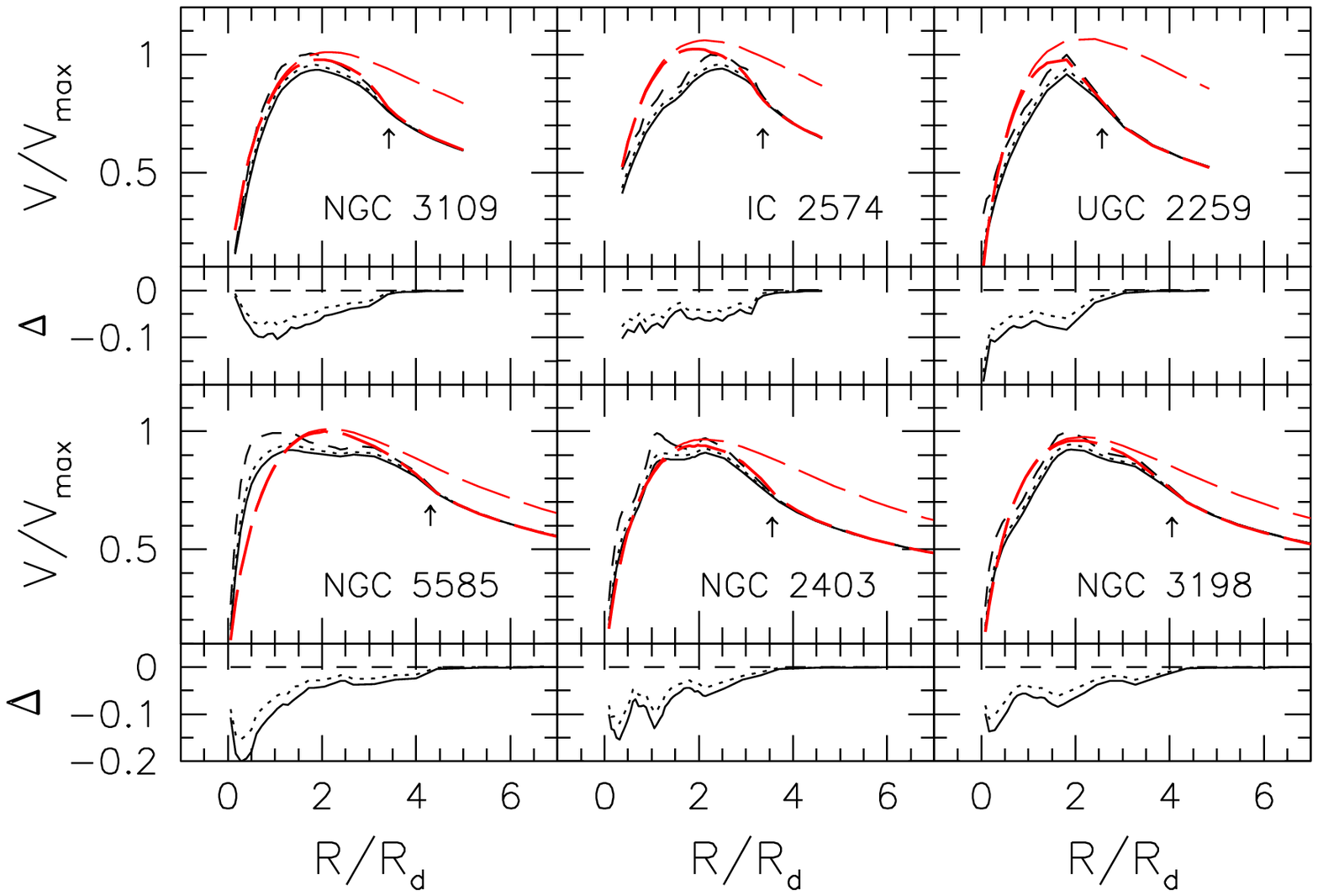}
\caption{Effect of disk thickness on disk circular velocity. 
The rotation curves were calculated from the observed surface brightness  
profile, with a disk thickness $q_{\rm d}=$ 0 ({\it dashed line}),
0.17 ({\it dotted line}), and 0.25 ({\it solid line}).  
For comparison we also show the rotation curves of exponential disks  
({\it long-dashed line}) truncated at the last surface brightness data
point and continued to infinity. 
The truncation radius is indicated by an arrow.  
The bottom panel shows the residuals with respect to the thin disk.} 
\end{center}
\end{figure*}

\section{Summary}\label{sec:discussion} 
There has been  much debate recently over the  shape of galaxy density
profiles, especially  regarding the center  of dwarf and  LSB galaxies
that  are believed  to be  dark  matter dominated.   However, a  large
number  of systematic  effects  such as  slit  position error,  poorly
sampled   velocity  fields,  and   non-circular  motions   thwart  the
straightforward interpretation of observed rotation curves as circular
velocity profiles.
 
Even  if the  circular  velocity profile  is  measured perfectly,  the
determination of $\alpha$ is  complicated by the degeneracies inherent
to the mass  modeling exercise.  The most cited  degeneracy is that of
the unknown value of  the stellar mass-to-light ratio, {$\Yd$}; strong
covariances  with the  halo  concentration and  density profile  slope
prevent  a definitive  determination  of $\Yd$.   Even  if $\Yd$  were
known, degeneracies that exist  between the halo parameters might also
prevent a unique determination of $\alpha$.
 
Independent constraints  may help in breaking  these degeneracies.  We
have  considered such  constraints with  six disk  galaxies  that have
\ha\,  and \hi\,  rotation  curves and  $R$-band  imaging.  The  \ha\,
rotation curves are  derived from two-dimensional Fabry-Perot velocity
fields (Blais-Ouellette 2000; Corradi \etal  1991) and as such are not
affected  by most  of  the  systematic caveats  raised  by Swaters  et
al.~(2003a)   on   the   context   of   long-slit   spectroscopy   and
low-resolution  radio-synthesis mapping, such  as slit  position error
and  beam smearing.   However, we  cannot exclude  the  possibility of
non-circular motion effects in these rotation curves.
 
The advantages of two-dimensional \ha\, velocity fields over long-slit
spectra are demonstrated with UGC  2259, also studied by Swaters \etal
(2003a).  Sampling more of the  velocity field yields a scatter in the
rotation  curve   data  of   Blais-Ouellette  \etal  (2004)   that  is
significantly smaller than that of  the long-slit spectrum of the same
object by  Swaters \etal (2003a).  These  authors find $\alpha=0.86\pm
0.18$ for a  minimum disk, and their plot  of $\chi^2$ versus $\alpha$
is mostly  flat for $0 < \alpha  < 1$ for all  $\Upsilon_{\rm d}$.  By
contrast,  we find the  best fit  $\alpha=0$ for  all $\YdR$,  and the
\chisqr increases  with $\alpha$.  It should be  noted that $\alpha=1$
halo fits  for UGC~2259 deviate most  strongly in the  central 0.5 kpc
($\simeq 0.01 R_{200}$) where systematic effects on the rotation curve
are most significant.
 
For galaxies  with appreciable  baryonic components, the  formation of
the disk may have altered the initial dark matter density profile.  To
first order the  halo contracts, but other mechanisms  such as stellar
feedback and stellar  bars may result in less  concentrated halos.  To
encompass  the full  range, we  run  fits with  and without  adiabatic
contraction.  Adiabatic  contraction has  the effect of  turning cores
into  cusps,  even for  relatively  low  mass  disks.  The  effect  of
adiabatic  contraction  on the  circular  velocity  and density  slope
increases  for  larger $\YdR$,  lower  $c_{-2}$,  and lower  $\alpha$.
Obviously, maximal disks are inconsistent with adiabatic contraction.
 
Applying the  SPS model of Bell \&  de Jong (2001) to  the expected or
observed $B-R$ colors for these galaxies implies that all galaxies are
sub-maximal at  2.2$R_d$. This is in agreement  with other independent
techniques  that suggest  that  HSB disk  galaxies  have, on  average,
sub-maximal  disks  with  $V_{\rm  disk}^{\rm  max}\simeq  0.6  V_{\rm
obs}^{\rm max}$ (e.g., Bottema 1993, 1997; Courteau \& Rix 1999; Trott
\& Webster 2002; Courteau \etal 2005).
 
In the  model from  $N$-body simulations of  Bullock \etal  (2001) for
halos  with $50 \lta  \v200 \lta  160 \kms$,  the $2-\sigma$  range in
concentrations  is   $6  \lta  \ccm2   \lta  30$.   By   defining  the
concentration parameter as  $c_{-2}=r_{200}/r_{-2}$, where $r_{-2}$ is
the  radius  where  the density  slope  of  the  halo is  $-2$,  these
constraints can be  applied to halos of arbitrary  $\alpha$.  All fits
have $\ccm2  \lta 30$, but  often fits with $\alpha\gta1$  have $\ccm2
\lta 6$. Applying this constraint to mass models thus lowers the range
of  acceptable   $\alpha$.   As   a  further  constraint,   we  impose
$V_{200}\le  V_{\rm  max}$;  this  constraint does  not  significantly
affect the  best fits, but it does  help to eliminate bad  ones. If we
impose the stronger constraint $V_{200}\le V_{\rm max}/1.4$ for bright
galaxies  (NGC 2403  and  NGC 3198),  values  of $\alpha  \gta 1$  are
disfavored.
 
Without constraints NGC 3109 and IC 2574 strongly favor $\alpha \simeq
0$ and  low $\YdR$,  although both of  these galaxies are  not ideally
suited  for   mass  modeling  studies  (NGC  3109   has  an  uncertain
inclination angle, and  IC 2574 has a disrupted  velocity field).  The
remaining four  galaxies are consistent  with a wide range  of central
density  slopes  $0\lta \alpha  \lta  1.4$  and mass-to-light  ratios,
$\YdR$.  Applying  our full  set of constraints  reduces the  range of
acceptable  $\alpha$,  but  taking  fits with  and  without  adiabatic
contraction  as two  extremes, there  still  remains a  wide range  of
acceptable $\alpha$, and only for NGC~5585 can we strongly distinguish
between  $\alpha\simeq0$  and   $\alpha\simeq1$  (in  this  case,  the
best-fit  $\alpha=0$).   Our best-fit  models  with constraints  favor
sub-maximal disk models, with $\vdisk/\vtot\lta 0.6$ at 2.2 disk scale
lengths for all six galaxies.
 
Accurately   determining  the   error  bars   (both   statistical  and
systematic) on  the observed rotation curve(s) is  crucial to breaking
the degeneracies.   Doubling the  minimum rotation curve  error values
from, say, 2 to just  4\kms reduces the differences in \chisqr between
models   with   different   $\alpha$   or  $\YdR$   to   statistically
insignificant levels.
 
Changing  the distance  or disk  thickness  of the  galaxy alters  the
best-fitting \YdR, although the relative difference in goodness of fit
between  $\alpha=0$  and $\alpha=1$  halos  is  not significant.   The
effect of  halo flattening is  to decrease its concentration,  but the
effect on $\chi^2$ is practically unchanged.
 
Thus, given  the above uncertainties, we conclude  that rotation curve
mass modeling of  disk galaxies fails to provide  tight constraints on
the  central  density  slope  of dark  matter  halos.\footnote{Similar
limitations for the  mass modeling of dwarf and  LSB disk galaxies are
addressed  in Swaters  \etal (2003),  and for  elliptical  galaxies in
C\^ot\'e  \etal (2003).}  Constraints  on central  density slopes  are
possibly  strongest  in low-mass  galaxies,  especially LSB  galaxies,
provided that there  are no systematic errors in  the rotation curves.
Unfortunately,  at present, the  predictions of  numerical simulations
for these galaxy types are the weakest.
 
However, the  prospects for determining  the relative amounts  of dark
and visible  matter in disk  galaxies (e.g., beyond 2$\Rd$,  where the
rotation  curve becomes  flatter)  look more  promising provided  that
near-IR imaging and SPS models or velocity dispersion measurements are
available to constrain $\Yd$.
 
Acknowledgements:  We  would  like  to thank  Lauren  MacArthur,  Joel
Primack and Frank van den Bosch for helpful discussions, Matt Choptuik
for use of the VN cluster  at UBC, and the referee for useful comments
that led to a more condensed  and focused presentation.  S.  C. and C.
C.   acknowledge  financial  support  from the  National  Science  and
Engineering Council of  Canada.  This research has made  use of NASA's
Astrophysics Data  System Abstract Service,  as well as  the NASA/IPAC
Extragalactic Database (NED), which  is operated by the Jet Propulsion
Laboratory,  California Institute of  Technology, under  contract with
the National Aeronautics and Space Administration.


\end{document}